\documentclass[traditabstract]{aa}
\usepackage{hyperref}
\usepackage{txfonts}
\usepackage{natbib}
\usepackage{graphicx}

\begin{document}
\bibpunct{(}{)}{;}{a}{}{,} 
\def\sizex{16.0 cm} \def\bigx{10.0 cm} \def\smallerxsize{7.0 cm}
\def\smallxsize{10.0 cm} \def\smallysize{12.0 cm} 
\def\kab{\hbox{$K_{\rm AB}$}}
\def\ks{\hbox{$K_{\rm s}$}}
 \catcode`\@=11
\def\gsim{\ifmmode{\mathrel{\mathpalette\@versim>}}
    \else{$\mathrel{\mathpalette\@versim>}$}\fi}
\def\lsim{\ifmmode{\mathrel{\mathpalette\@versim<}}
    \else{$\mathrel{\mathpalette\@versim<}$}\fi}
\def\@versim#1#2{\lower 2.9truept \vbox{\baselineskip 0pt \lineskip 
    0.5truept \ialign{$\m@th#1\hfil##\hfil$\crcr#2\crcr\sim\crcr}}}
\catcode`\@=12
\def\advsr{Adv.\ Space Res.}%
\newcommand\fnurl[2]{%
  \href{#2}{#1}\footnote{\url{#2}}%
}

\title {UltraVISTA: a new ultra-deep near-infrared survey in COSMOS\thanks{Based on data products from observations made with ESO
    Telescopes at the La Silla Paranal Observatory under ESO programme
    ID 179.A-2005 and on data products produced by TERAPIX and the
    Cambridge Astronomy Survey Unit on behalf of the UltraVISTA
    consortium.}\fnmsep{}\thanks{Catalogs are available at the CDS via
    anonymous ftp to cdsarc.u-strasbg.fr (130.79.128.5) or from
    \url{http://cdsarc.u-strasbg.fr/viz-bin/qcat?J/A+A/544/A156}}} 

\offprints {H.~J.~McCracken} \date{Received 30 April 2012 /
  Accepted 22 June 2012} \titlerunning{first UltraVISTA data release} 
\authorrunning{McCracken et al.}  
\author{H. J. McCracken\inst{1}, B. Milvang-Jensen\inst{2},
  J. Dunlop\inst{3}, M. Franx\inst{4}, J.~P.~U. Fynbo\inst{2}, O. le
  F\`evre\inst{5}\\J. Holt\inst{4}, K.~I. Caputi\inst{3,6},
  Y. Goranova\inst{1}, F. Buitrago\inst{3}, J.~P. Emerson \inst{7},
  W. Freudling\inst{8}, P. Hudelot\inst{1},\\
  C.~L\'opez-Sanjuan\inst{5}, F. Magnard\inst{1}, Y. Mellier\inst{1}, P. M{\o}ller\inst{8},
  K.~K.  Nilsson\inst{2}, W. Sutherland\inst{7},\\ L. Tasca\inst{5} \and J. Zabl\inst{2}}

\institute{ Institut d'Astrophysique de Paris, UMR7095 CNRS,
  Universit\'e Pierre et Marie Curie, 98 bis Boulevard Arago, 75014
  Paris, France \and Dark Cosmology Centre, Niels Bohr Institute,
  University of Copenhagen, Juliane Maries Vej 30, 2100 Copenhagen,
  Denmark \and SUPA, Institute for Astronomy, University of Edinburgh,
  Royal Observatory, Edinburgh EH9 3HJL, UK\and Leiden Observatory, Leiden University,
  P. O. Box 9513, NL-2300 RA Leiden, The Netherlands\and Laboratoire
  d'Astrophysique de Marseille, CNRS and Aix-Marseille Universit\'e, 38
  rue Fr\'ed\'eric Joliot-Curie, 13388 Marseille Cedex 13, France\and Kapteyn
  Astronomical Institute, University of Groningen, P.O. Box 800, 9700
AV Groningen, The Netherlands \and Astronomy
  Unit, School of Physics and Astronomy, Queen Mary University of
  London, Mile End Road, London, E1 4NS, UK\and European Southern Observatory, Karl-Schwarzschild-Strasse 2,85748 Garching bei M{\"u}nchen, Germany}

\abstract{In this paper we describe the first data release of the
  UltraVISTA near-infrared imaging survey of the COSMOS field. We
  summarise the key goals and design of the survey and provide a
  detailed description of our data reduction techniques. We provide
  stacked, sky-subtracted images in $YJHK_{\rm s}$ and narrow-band
  filters constructed from data collected during the first year of
  UltraVISTA observations. Our stacked images reach $5\sigma$ $AB$
  depths in an aperture of $2\arcsec$ diameter of $\sim 25$ in $Y$ and
  $\sim 24$ in $JHK_{\rm s}$ bands and all have sub-arcsecond
  seeing. To this $5\sigma$ limit, our $K_{\rm s}$ catalogue contains
  216,268 sources. We carry out a series of quality assessment tests
  on our images and catalogues, comparing our stacks with existing
  catalogues. The $1\sigma$ astrometric RMS in both directions for
  stars selected with $17.0<K_{\rm s}\rm {(AB)} <19.5$ is $\sim
  0.08\arcsec$ in comparison to the publicly-available COSMOS ACS
  catalogues. Our images are resampled to the same pixel scale and
  tangent point as the publicly available COSMOS data and so may be
  easily used to generate multi-colour catalogues using this data. All
  images and catalogues presented in this paper are publicly available
  through ESO's ``phase 3'' archiving and distribution system and from
  the UltraVISTA web site.
   }
\authorrunning{H.J.~McCracken et al.}
\maketitle

\section{Introduction}
\label{sec:intro}

The vital role of near-infrared ($\lambda \simeq 1 - 2.5\,\mu$m)
imaging surveys for advancing our understanding of galaxy evolution
has long been recognised
\citep{Cowie:1990p9035,Glazebrook:1991p12757}. 
While optical surveys utilising large-format charge-coupled device
(CCD) detectors were the first to enable the discovery of substantial
samples of normal galaxies at redshifts $z > 2$
\citep{Steidel:1996p9040,Madau:1996p12058}, it was already known that
at least some galaxies at high redshift were either too old or too
dust-obscured to be easily detected by rest-frame near-ultraviolet
selection \citep{Dunlop:1996p12783,Dey:1996p12819}. In addition, even
for apparently young UV-luminous galaxies, the value of using
near-infrared observations to sample the rest-frame optical light,
more representative of the evolved mass-dominant stellar population,
was understood and indeed demonstrated before the advent of
multi-pixel near-infrared imagers \citep{Lilly:1984p12821}.

However, near-infrared surveys are a challenging proposition for
several reasons. Firstly, at near-infrared bandpasses the sky
background is extremely bright; in the $K_{\rm s}$ band, in $AB$ magnitudes,
it is typically $15~\rm {mag}/\rm arcsec^2$, which means that many
short exposures must be combined in order to avoid detector saturation
on the sky, greatly increasing overheads. Secondly, the sky background
is time-variable, many magnitudes brighter than the faint astronomical
sources of interest, and so must be carefully subtracted from each
image before scientific exploitation can take place. Lastly,
conventional silicon CCDs are very inefficient at near-infrared
wavelengths, and a different detector technology must be employed
which is an order of magnitude more expensive. In terms of sky
footprint, near-infrared detectors have generally lagged behind
optical detectors by approximately a decade.

Nevertheless, these challenges have been progressively overcome, and,
following the pioneering work described above with early near-infrared
arrays such as IRCAM on the UK Infrared Telescope
(UKIRT) \citep{McLean:1986p13369}, the full potential of near-infrared
surveys to clarify our view of galaxy evolution at $z \simeq 1 - 3$
began to be realised with the advent of larger format infrared array
cameras such as ISAAC on ESO's Very Large Telescope (VLT)
\citep{Cimatti:2002p3013,Labbe:2003p8709,Franx:2003p310}. Meanwhile,
the importance of near-infrared surveys for revealing dust-obscured
star-forming galaxies was further enhanced by the discovery of
significant numbers of dusty-enshrouded high-redshift star-forming
galaxies at sub-mm wavelengths
\citep{Hughes:1998p13731,Scott:2002p13265}. Around the same time the
unique power of the deepest near-infrared imaging to conduct
rest-frame ultraviolet surveys for galaxies at $z > 6.5$ was first
demonstrated using the NICMOS camera on the Hubble Space Telescope
\citep[HST;][]{Bouwens:2004p12918,Thompson:2005p12949}

Despite these impressive advances, the field-of-view offered by
near-infrared cameras such as IRCAM, NICMOS and ISAAC was very small
(a few arcmin$^2$), and it is only in the last half-decade or so that
the introduction of genuinely large-format near-infrared array cameras
has enabled efficient, deep near-infrared imaging of degree-scale
areas of sky, allowing studies of more representative volumes of the
high-redshift universe (i.e. $\simeq 100 \times 100$ comoving
Mpc). First WFCAM on UKIRT \citep{Casali:2007p13430}, then WIRCam on
the Canada France Hawaii Telescope (CFHT) \citep{Puget:2004p4595}, and
NEWFIRM at NOAO \citep{Probst:2004p13554} have heralded a new era of
major coordinated near-infrared survey programmes (e.g. UKIDSS,
\cite{Lawrence:2007p13267}; NEWFIRM Medium-Band Survey,
\cite{vanDokkum:2009p13556}); WIRDS and associated near-infrared
follow-up of the COSMOS field
\citep{Bielby:2011p12309,McCracken:2010p10723}. This has led to a
number of breakthroughs in extra-galactic astronomy, including, for
example, the study of the bright end of the galaxy luminosity function
from $z = 0$ out to $z \simeq 6$
\citep{McLure:2009p13328,Cirasuolo:2010p13329}, and the discovery of
the most distant known quasar \citep{Mortlock:2011p13333}.

In addition, deep, wide-field near-infrared photometry coupled with
high-quality optical surveys has enabled spectral energy distribution
(SED) fitting techniques to be pushed beyond $z\sim1.5$. Near-infrared
data play a key role in minimising the catastrophic failure rates in
photometric redshift estimates and provides robust rest-frame visible
flux determinations at $z\sim2$ \citep{Ilbert:2009p4457}, enabling
measurements of the evolution of the mass buildup in stars over a
large fraction of the age of the Universe
\citep{Drory:2005p8438,Arnouts:2007p3665,Ilbert:2010p10744,Caputi:2011p13697}.

These efforts have now culminated in VISTA \citep{2010SPIE.7733E...4E}
the first 4-m class telescope specifically designed to conduct
wide-area near-infrared surveys and equipped with a large-format array
camera, ``VIRCAM'' \citep{Dalton:2006p12606}. Thanks to its large
mosaic of 16 detectors, VIRCAM is currently the most efficient
wide-field near-infrared survey camera in the world (around four times
more efficient than WIRCam, and three times as efficient as WFCAM). It
also has the benefit of being mounted on a telescope for which
virtually all observing time is available for surveys, and for which
observations are efficiently programmed in queue-scheduled mode.
Inspired by the success of UKIDSS, ESO has implemented a coordinated
multi-tier public survey programme with VISTA. The UltraVISTA survey
presented here is the deepest component of the VISTA survey ``wedding
cake''.

Covering an area of 1.5\,deg$^2$, UltraVISTA is significantly larger
than the only comparably-deep near-infrared survey conducted to date
(the UKIDSS Ultra Deep Survey\citep[UDS;][]{Almaini:2007p13557}), and
will ultimately go significantly deeper. VIRCAM also offers two
significant advantages over WFCAM (and indeed WIRCam or NEWFIRM) in
that its Raytheon detectors are much more sensitive in $Y$-band, and
are essentially free from the electronic cross-talk. These are crucial
benefits in the planned exploitation of UltraVISTA for the discovery
of the most luminous galaxies at $z \simeq 7$,
e.g., \cite{Anonymous:tylwsel8}. 

To maximise the leverage and legacy value of these new deep
near-infrared data, the UltraVISTA survey is centred on the COSMOS
field, the location of the largest ever ACS optical mosaic obtained
with HST \citep{Scoville:2007p13732,2007ApJS..172..196K} and an ever
growing heritage of deep \fnurl{ground-based and space-based
  multi-frequency imaging and
  spectroscopy}{http://cosmos.astro.caltech.edu}. The first-year data
set described in this paper is already deeper than all existing COSMOS
NIR data \citep{McCracken:2010p10723,Bielby:2011p12309} in all bands
by between one and two magnitudes and also contains for the first time
deep $Y-$ band imaging.

To most efficiently exploit VISTA for the discovery and study of
UV-selected galaxies at the highest redshifts ($z \simeq 6.5 - 9$)
{\it and} in the investigation of the growth of galaxies through the
crucial redshift range $1 < z < 3$ when cosmic star-formation density
peaks \citep{Hopkins:2006p13039}, the UltraVISTA survey comprises
three separate components: a wide, deep $Y,J,H,K_{\rm s}$ survey (a
contiguous field covering $\simeq 1.5$\,deg$^2$); an ultra-deep
$Y,J,H,K_s$ survey (consisting of deeper strips covering $\simeq
0.7$\,deg$^2$), and an ultra-deep narrow-band ($\lambda =
1.18$\,$\mu$m) survey targeting emission-line galaxies at a range of
redshifts, e.g. H$\alpha$ at $z=0.8$, [OIII]-emitters at
$z=1.4$, [OII] emitters at $z=2.2$, and ultimately Ly$\alpha$ emitters
at $z=8.8$. To accomplish these goals, UltraVISTA has been allocated
1800hrs of execution time.

It is important to stress that while the advent of Wide Field Camera 3
(WFC3) on HST in 2009 has enabled extremely deep near-infrared imaging
(up to $\lambda \simeq 1.6$\,$\mu$m) which has revolutionised the
study of galaxies at $z \simeq 7 - 8$,
\citep{Bouwens:2010p13076,McLure:2010p13065,Oesch:2010p13068,Finkelstein:2010p13233,Bunker:2010p13224}
the very small field-of-view offered by WFC3/IR coupled with its
inability to observe in the $K$-band means that deep ground-based
surveys such as UltraVISTA remain of crucial importance. In
particular, the largest current (or indeed planned) WFC3/IR
extragalactic survey is the Cosmic Assembly Near-infrared Deep
Extragalactic Legacy Survey (CANDELS;
\cite{Grogin:2011p13008,Koekemoer:2011p12718}), but even this
900-orbit 3-year HST Treasury Program will only cover $\simeq
800$\,arcmin$^2$. Thus UltraVISTA is an excellent complement to
CANDELS, and indeed CANDELS has recently completed deep $J,H$-band
WFC3/IR imaging of a $\simeq 200$\,arcmin$^2$ region within the
1.5\,deg$^2$ UltraVISTA imaging described here (i.e. covering only
$\simeq 4$\% of UltraVISTA).

In this paper we present a detailed description of the data reduction
methods and properties of the five near-infrared stacks created from
the first season of UltraVISTA operations. Already, with only these
first images, the UltraVISTA survey has the largest \textit{\'etendue}
of \textit{any} near-infrared survey.

All magnitudes in this paper, unless otherwise noted, are given in the
AB system. Data products described here are available from
\fnurl{ESO}{http://www.eso.org/sci/observing/phase3/data_releases.html},
\fnurl{the UltraVISTA website}{http://www.ultravista.org/} and
\fnurl{CESAM}{http://cesam.oamp.fr/ultravista/index.php}. 

\section{Observations and data reductions}
\label{sec:observ-data-reduct}

\subsection{Observations}
\label{sec:observations}

The images described here were taken between 5th December 2009 and the
19th of April 2010 with the VIRCAM instrument on the VISTA telescope
at Paranal as part of the UltraVISTA survey program. VIRCAM is a
wide-field near-infrared camera consisting of 16 $2048\times2048$
Raytheon VIRGO HgCdTe arrays arranged in a sparse-filled array with
gaps between each array of 0.90 \& 0.425 of a detector in X and Y
respectively \citep{2010SPIE.7733E...4E}. The mean pixel scale
is $0.34\arcsec {\rm pixel}^{-1}$
\citep{Dalton:2006p12606}. 

The sky coverage of the 16 non-contiguous detectors is called a
``pawprint''.  A contiguous region of size $1.5^\circ \times
1.23^\circ$ can be covered by means of six pawprints suitably spaced
in right ascension and declination with random $60\arcsec$ jitter
offsets in both directions (two $\approx 0.1^\circ$ bands at the top
and bottom of the field receive half the exposure time).

Specifically, three pawprints with identical RA and with Dec differing
by 5.5$'$ = 47.5\% of a detector height make up a set of four stripes
(corresponding to the ultra-deep stripes in UltraVISTA), and another
three pawprints shifted by 95\% of a detector width in RA make up
another set of stripes, which together form a contiguous region where
most pixels in the resulting stack are covered by two of the six
pawprints.

\begin{figure}
       \includegraphics[width=0.49\textwidth]{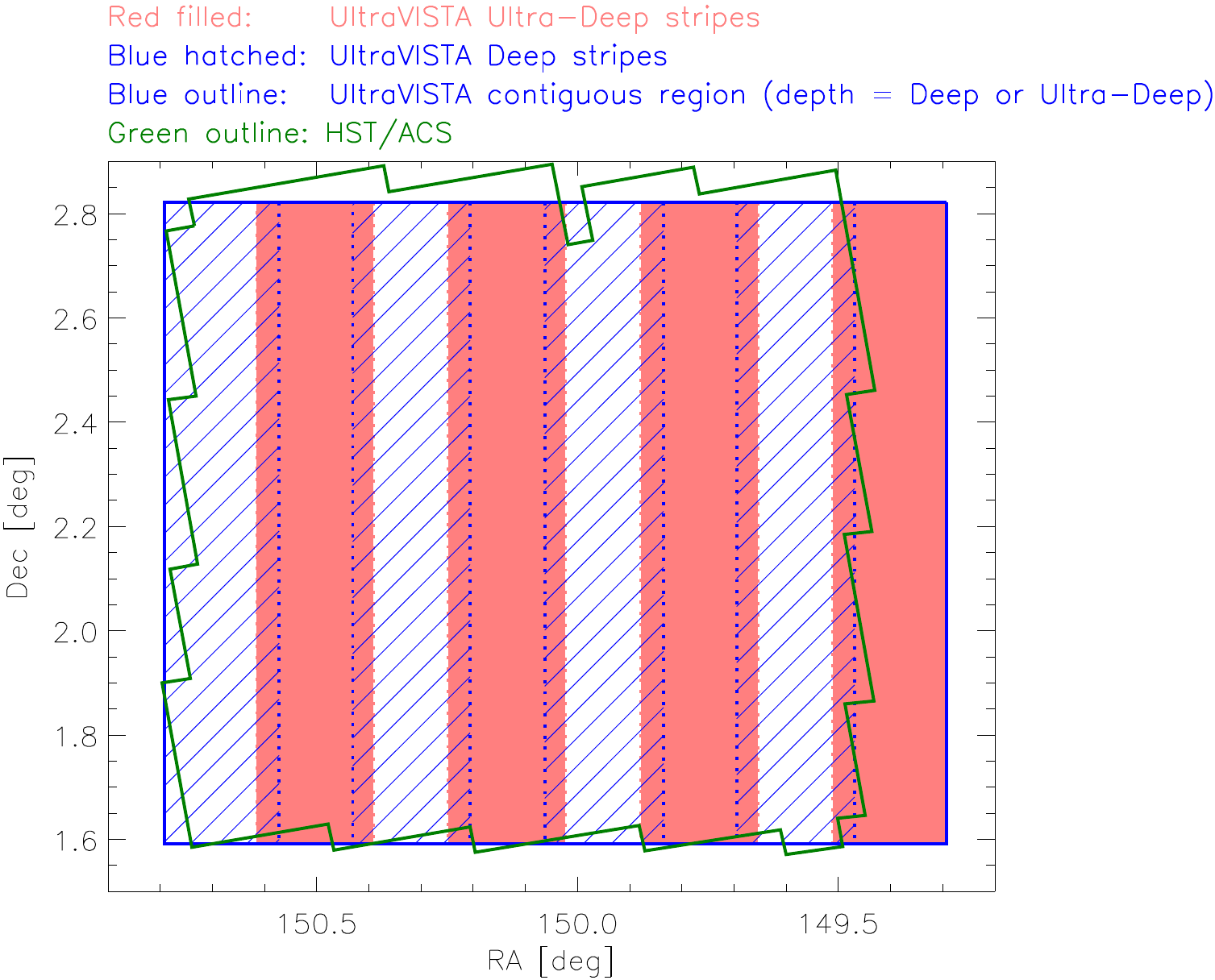}
       \caption{Schematic layout of UltraVISTA observations, showing
         deep and ultra-deep regions (hatched and filled regions
         respectively). The data described in this paper correspond to
         a uniform coverage in YJHKs of the contiguous region and to
         NB118 observations of the ultra-deep stripes.}
   \label{uvista-layout}
 \end{figure}

Fig.~\ref{uvista-layout} illustrates the layout of UltraVISTA
observations showing the deep survey, which will cover the full survey
area, and the ultra-deep part, which covers half of this area in a
series of ultra-deep stripes. The first season of UltraVISTA data
described in this paper comprises six contiguous pawprints in four
broad-band filters covering the deep survey area, each with equal
exposure times, and narrow band observations on the ultra-deep
stripes; subsequent observing seasons are expected to concentrate
exclusively on the ultra-deep stripes.

\begin{table*}
\caption{Characteristics of the OBs used in UltraVISTA season 1}
\label{tab:OBinfo}
\begin{center}
\renewcommand{\arraystretch}{1.00} 
\setlength{\tabcolsep}{4pt} 
\begin{tabular}{llrrccrcccr}
\hline
\hline
OB description & Filter & DIT [s] & NDIT & \multicolumn{3}{c}{Jitter parameters}               & Nesting & N$_\mathrm{pawprints}$ & Total exp.\ [s] & N$_\mathrm{OBs}$ \\
               &        &         &      & Pattern        & Amplitude [$''$]  & N$_\mathrm{jit}$ &         &              &                         &                  \\
\hline                                                          
$Y$                & $Y$      &  30     &  4   & Random  & 60               & 30    & FPJME   & 1           & 3600            & 36    \\
$J$                & $J$      &  30     &  4   & Random  & 60               & 30    & FPJME   & 1           & 3600            & 37    \\
$H$                & $H$      &   6     & 10   & Random  & 60               & 60    & FPJME   & 1           & 3600            & 36    \\
$K_{\rm s}$ long OB       & $K_{\rm s}$     &  10     &  6   & Random  & 60               & 60    & FPJME   & 1           & 3600            & 18    \\
$K_{\rm s}$ short OB      & $K_{\rm s}$    &  10     &  6   & Random  & 60               & 30    & FPJME   & 1           & 1800            & 27    \\
NB118 single paw & NB118  & 300     &  1   & Random  & 61               & 11    & FPJME   & 1           & 3300            &  6    \\
NB118 three paws & NB118  & 280     &  1   & Random  & 61               &  4    & FJPME   & 3           & 3360            &  4    \\
\hline
\end{tabular}
\end{center}

\vspace*{0.1ex}

Notes --
The ``Amplitude'' column gives the Maximum Jitter Amplitude, where a value
of 60$''$ corresponds to jitter positions being drawn from a random, uniform
distribution over a box of side length 120$''$, centered on the nominal
centre coordinates of the given pawprint.
The ``Nesting'' column indicates the order in which different operations
are done, see text.
The ``N$_\mathrm{pawprints}$'' column gives
the number of pawprints done by the given OB\@.
To cover the contiguous UltraVISTA field in an approximately uniform manner,
a set of 6 OBs of type N$_\mathrm{pawprints}$ = 1 are needed, each centered
on the pawprint in question.
The ``Total exp.'' column gives the total exposure time contained in the OB;
this number is DIT$\times$NDIT$\times$N$_\mathrm{jit}$$\times$N$_\mathrm{pawprints}$.
The ``N$_\mathrm{OBs}$'' column gives the number of OBs of the given type that
are associated with the data covered by this paper.
\end{table*}

The observations, carried out in service mode, are specified by
observation blocks (OBs).  The characteristics of the OBs used in
UltraVISTA season one are listed in Table~\ref{tab:OBinfo}. Most of
the season one OBs comprise images jittered around the centre of a
single pawprint position, with the jitters being drawn from a random,
uniform distribution over a box of side length 120$''$ (random jitters
are necessary because of persistence effects in VIRCAM and are also
essential to derive a good sky frame). 

The exception to this was the ``NB118 three paws'' OBs
(Table~\ref{tab:OBinfo}, which comprised images jittered around the
centres of the \emph{three} pawprints forming the ultra-deep stripes.
For OBs containing more than a single pawprint per OB, the nesting
(Table~\ref{tab:OBinfo}) is important, and we did not use the optimal
value.  These OBs had a nesting of ``FJPME'' such that F (filter) is
the outermost loop, and E (expose) is the innermost loop. The
important aspect here is that the three pawprints (P) (spaced exactly
by 5.5$'$ in Dec) are completed before a random jitter (J) is
applied. This means that the faint persistent images (i.e.\ fake
sources that are memories of a bright star at that x,y position on the
detector in the one or two previous exposures) will be present in the
stack at positions located 5.5$'$ (and 11$'$) away from bright stars
in DEC. We deal with this by masking the persistent images in the
individual NB118 images (see Milvang-Jensen et al., in prep.\ for
details of the procedure). For the other UltraVISTA OBs, the faint
persistent images are fully removed by the sigma clipping used in
producing the stacks, thanks to the random jitters applied between
each single exposure. The first season of observations described here
comprise around 200 OBs in total. The average efficiency (calculated
as the total exposure time divided by total execution time these OBs)
was 77\%.

In light of our experience gained in the season one observations
described here, from season 2 onwards we modified some of the OBs. For
$Y$, we changed the DIT to 60 sec (with NDIT = 2), since 30 sec was
unnecessarily short; for $H$, we changed the DIT to 10 sec (with NDIT
= 6), for the same reason.  For NB118, we changed the DIT to 120 sec
(with NDIT = 1), since 300 sec was unnecessarily long. We also changed
our observation strategy to jitters centered around a single
pawprint per OB, and changed the total exposure time per OB to 1 hour
(corresponding to 30 jittered exposures in an OB).

\subsection{Image selection and grading}
\label{sec:pre-reductions}

 \begin{figure*}
   \begin{center}
     \begin{tabular}{c@{}c@{}}
       \includegraphics[width=0.49\textwidth]{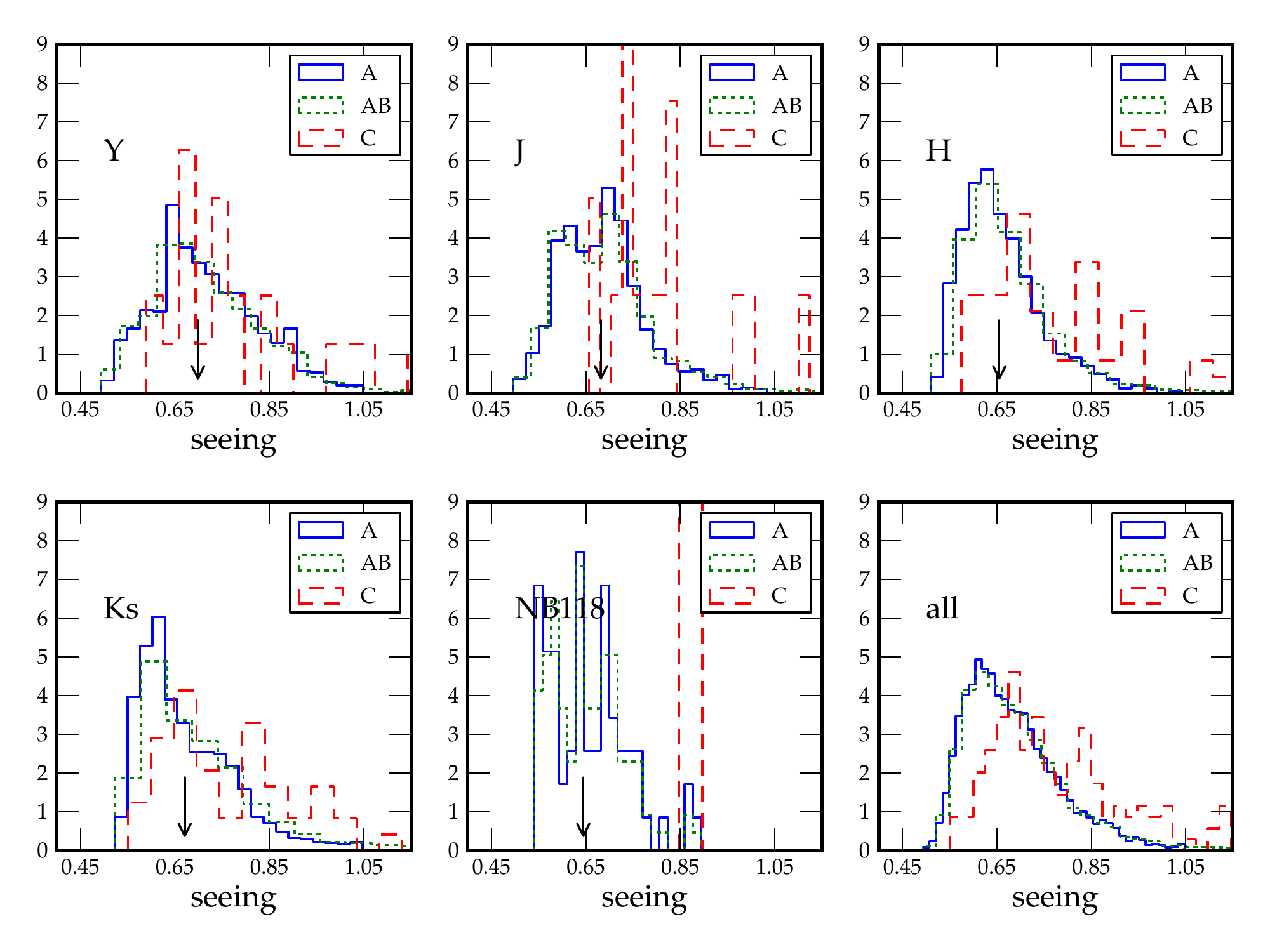}    &
       \includegraphics[width=0.49\textwidth]{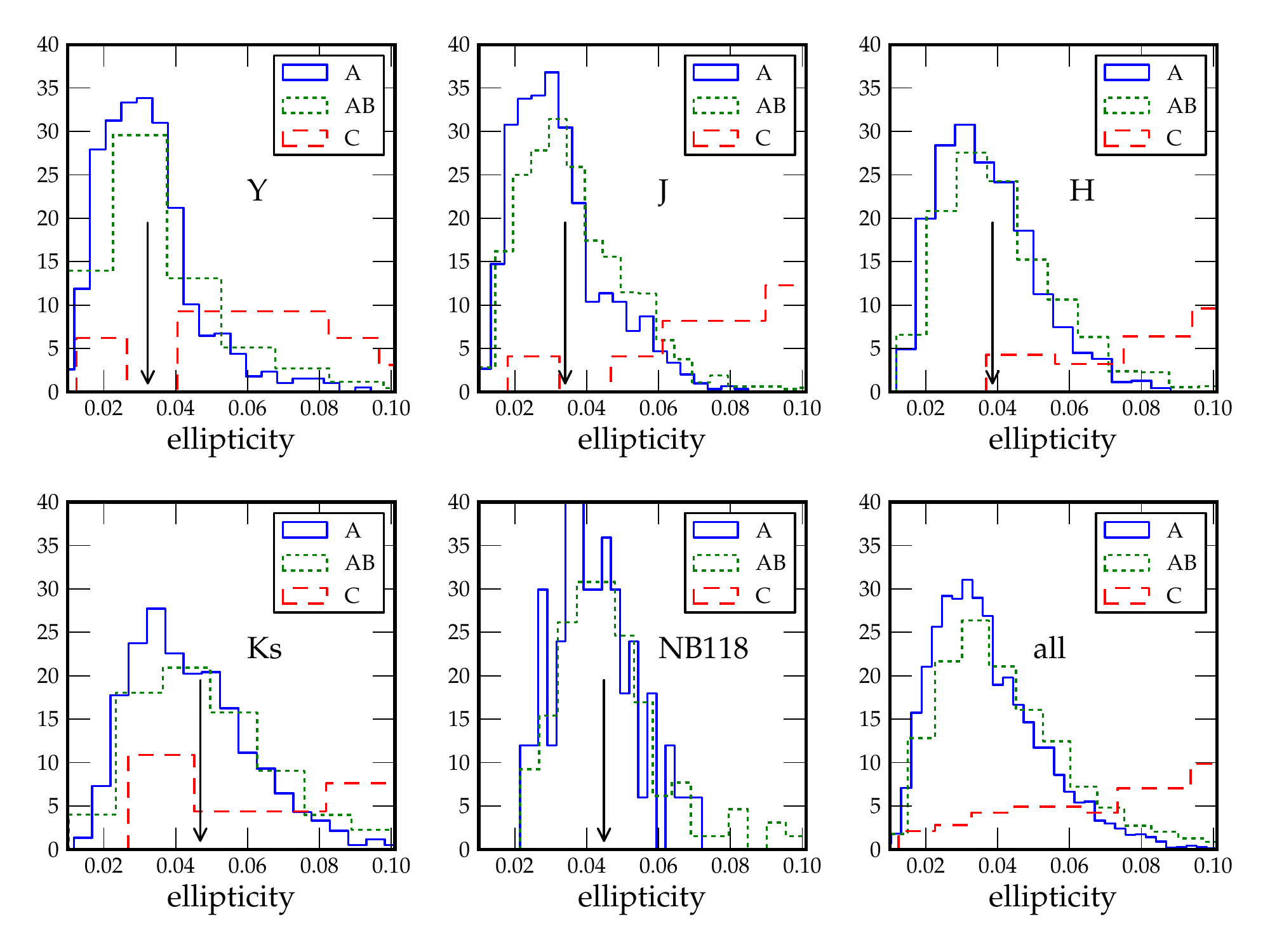}\\
     \end{tabular}
   \end{center}
   \caption{Seeing (left) and ellipticity (right) distributions for
     all UltraVISTA images considered. The arrow represents the median
     of each distribution for images classified as A or AB. Note that
     the distributions for each quality class have been re-normalised,
     and the vertical axis has been rescaled. A, AB, C represent the
     quality classifications described in the text.}
   \label{ellipticty_maps}
 \end{figure*}

 VIRCAM images are transferred to the Cambridge Astronomy Survey Unit
 \fnurl{(CASU)}{http://casu.ast.cam.ac.uk/surveys-projects/vista/technical/data-processing}
 for pre-preprocessing and removal of the instrumental signature. This
 includes dark subtraction, correction for rest anomaly,
 flat-fielding, initial sky-subtraction, de-striping, non-linearity
 corrections and gain normalisation\citep{Irwin:2004p13800}. CASU
 subsequently provides these pre-processed images for each survey, as
 well as stacks of images from a single OB and pawprint, comprising
 typically 30 or 60 images.

 For UltraVISTA we start from the individual pre-processed images,
 rather than the stacked OB blocks, for a number of reasons: firstly,
 the OB blocks are combined at CASU at the native pixel scale of the
 instrument, which means that in good seeing conditions (median FWHM
 $\sim 0.6\arcsec$) VIRCAM data is under-sampled. For this reason it
 is preferable to re-sample these data at a finer pixel scale;
 secondly, one of the principal scientific aims of the UltraVISTA
 project is to make measurements of distant ($z>6$) and faint ($K_{\rm
   s}\simeq 24$) galaxies. To do this requires extremely accurate
 removal of the sky background for each individual image; in the
 version of the CASU pipeline we used, a single sky background was
 used for all images coming from a given OB, and objects were not
 masked using the deepest possible mask. Given that the sky background
 is known to vary on shorter timescales, this process may lead to a
 systematic magnitude offset at faint magnitudes near bright
 sources. For these reasons we use an iterative sky-background removal
 technique starting from the pre-processed images and also resample
 all data to a pixel scale of $0.15\arcsec{\rm pixel}^{^-1}$.

 The images in this release were taken between 5th December 2009 and
 the 19th of April 2010. This does not consist of the complete number
 of images taken for the UltraVISTA program in the 2009-2010 observing
 season; subsequently, around 10\% additional images in H and Ks were
 made available by CASU using a different pipeline processing, after
 we had already graded the first batch of images. In order to maintain
 as a homogenous as possible data set, we restrict in this release
 ourselves to this initial batch. However, had we included these data,
 the average exposure time per pixel would have been 4800 s and 5400 s
 higher in $H$ and $K_{\rm s}$ respectively, i.e. only around 10\%
 larger. The images considered here were all processed with v0.8 of
 the CASU pre-processing pipeline and in total, we consider 7031
 individual images (each of which is a single multi-extension fits
 image containing 16 image extensions one for each of the VISTA
 detectors).

 Since UltraVISTA represents the first significant amount of data from
 VIRCAM processed at TERAPIX, we wished to visually inspect all images
 to identify any problems which had been potentially overlooked by the
 automatic pipelines. Therefore, all images were inspected and graded
 in the YOUPI\footnote{\texttt{http://youpi.terapix.fr/}}
 environment. Images were assigned a grade of A , B (usable for
 science), C, D (rejected). The left and right panels of
 Fig.~\ref{ellipticty_maps} shows the seeing FWHM (measured assuming a
 Gaussian core), ellipticity and grading distributions for all
 images. Based on these distributions we decided to keep all images
 which have stellar FWHM $<1.0 \arcsec$ and ellipticity $< 0.1$ and
 which were classified as either A or B based on visual
 inspection. The visual inspection process in general finds images
 which have bad PSFs or other optical defects which would have not
 been found by a typical seeing or ellipticity cut\footnote{Some of
   these bad PSFs were caused in part by software errors in early
   versions of ESO's Survey Area Definition tool: all season 1 OBs had
   pointing centres such that when a jitter jump went too far in one
   direction, the guide star fell outside the guide CCD; guiding was
   not active for the remaining images of that OB. This was fixed in
   season two observations by moving the pawprint centers.  These
   tracking errors produce double-lobed PSFs in some images; each of
   the \textit{individual} PSFs are smaller than the requirement and
   so pass our cut.}. In total we reject 426 images or around $6\%$ of
 the total.

 We do not use the confidence maps provided by CASU, but create our
 own weight maps from the supplied flat-fields and bad pixel maps
 using the \texttt{weightwatcher} tool \citep{Marmo:2008p13562}. For
 our NB118 images which were taken at a fixed set of jitter patterns
 and thus suffer from image persistence effects, we mask the
 persistent images using the procedure described in Milvang-Jensen et
 al. (in prep.).

\subsection{Two-step sky subtraction}
\label{sec:sky-subtraction}

To derive our sky-subtracted images, we use a set of tools developed
at TERAPIX which run under the distributed processing environment
``condor''\footnote{\texttt{http://www.cs.wisc.edu/condor/}}. (These
processing steps are described fully in
\cite{Bielby:2011p12309}). Sky-subtraction is a two-step iterative
process. To summarise, we start by adding back the sky background
frames subtracted by CASU (which are supplied as part of the original
data release.) Based on the first-pass stack (computed using the CASU
sky-subtracted images) and astrometric solutions, we compute object
masks for each individual image. Next, we use these object masks
(appropriately resampled based on an initial astrometric solution) to
effectively remove objects computed from a running sky for each
individual image, based on a median of images taken during a 20-minute
sliding window. After the subtraction of the running sky, we
re-''destripe'' the images and remove large-scale background gradients
using \texttt{sextractor} \citep{Bertin:1996p13615}. In general,
computing sky backgrounds for each of the 7000 images is highly
processor intensive; for each image, it takes around 15-20 minutes on
a standard TERAPIX computing node.

\subsection{Astrometric and photometric solutions}
\label{sec:astr-solut}

After sky-subtraction, weight maps and catalogues are computed once
more for each image using \texttt{QualityFITS}. Saturated objects,
based on an examination of the distribution of objects in the peak
surface brightness / magnitude plane, are flagged in these catalogues,
and the weight maps are used to flag cosmic rays. Bad pixels are also
flagged. Next, these catalogues are used to compute the final
astrometric and photometric solutions which will be used to combine
and scale the images. Astrometric solutions are computed independently
from each filter using the \texttt{scamp} tool
\citep{Bertin:2006p13607}, but use a common astrometric reference
catalogue drawn from the COSMOS $i-$band CFHT data (the same reference
catalogue used in \cite{Capak:2007p267} and
\cite{McCracken:2010p10723}). We use a third-order polynomial solution
in $x$ and $y$ detector co-ordinates (note that unlike the CASU
reductions, we do not assume a radially symmetric astrometric
solution). In order to derive a more robust astrometric solution, we
use a precomputed ``\texttt{.ahead}'' file for all images which
specifies the relative positions and orientations of each of the
sixteen detectors. In addition, we require that all the detectors
share a common tangent point (focal plane mode
``\texttt{SAME\_CRVAL}'' in \texttt{scamp}). These steps ensure that
we can reliably match our reference astrometric catalogues for many
thousands of input images (note that we do not use the higher-order
terms of the initial astrometric solution provided by
\fnurl{CASU}{http://casu.ast.cam.ac.uk/surveys-projects/vista/technical/astrometric-properties}). Thanks
to our densely sampled astrometric reference catalogue the internal
sigma of our astrometric solution is $\sim0.08\arcsec,0.09\arcsec$ in
directions North--South, East--West directions respectively.

Compared to our reference catalogue, in the same directions, we find
standard deviations of $\sim0.09\arcsec,0.10\arcsec$. Similar
values are found in all filters. Given that the native pixel scale of
VIRCAM is $0.34\arcsec\rm pixel^{-1}$, our astrometric solution is
more than sufficient to provide a precise and reliable image
coaddition (in fact, our astrometric accuracy is probably limited by
undersampling in the VIRCAM images).

Our initial magnitude zero points for each individual image are based
on those supplied by CASU for their \textit{.st} stacks (which
comprise a stack of several individual images), which is based on
their calibration of the VISTA photometric system's zero points. To
account for possible photometric variations between the images in each
\textit{.st} stack we calculate a rescaling factor for each using
\texttt{scamp} based on overlapping paw-prints. Note that the same
rescaling factors are applied to all detectors: we assume that the
relative scaling factors between chips does not change (the CASU
processing pipeline equalises the gain between all detectors at the
flat-fielding stage, and should remain constant). To create our final
stacks in the AB magnitude system \citep{Oke:1974p12716} we simply
apply the appropriate flux scaling to convert the supplied Vega
magnitudes to AB, based on the VISTA telescope detector, filter and
atmosphere combination. The conversion factor C from AB to Vega we use
are as follows, in the sense $\rm{ mag}_{\rm AB}={\rm mag}_{\rm
  vega}+C$ where $C=0.61, 0.90, 1.38, 1.84, 0.86$ for $Y, J, H, K_{\rm
  s}$ and NB118 filters respectively.

Note that CASU produces ``flat'' images which have constant flux per
pixel for a uniform illumination; this is taken into account in the
resampling stage.

\subsection{Coadded images}
\label{sec:stack-data-prod}

 \begin{figure*}
   \begin{center}
       \includegraphics[width=1\textwidth]{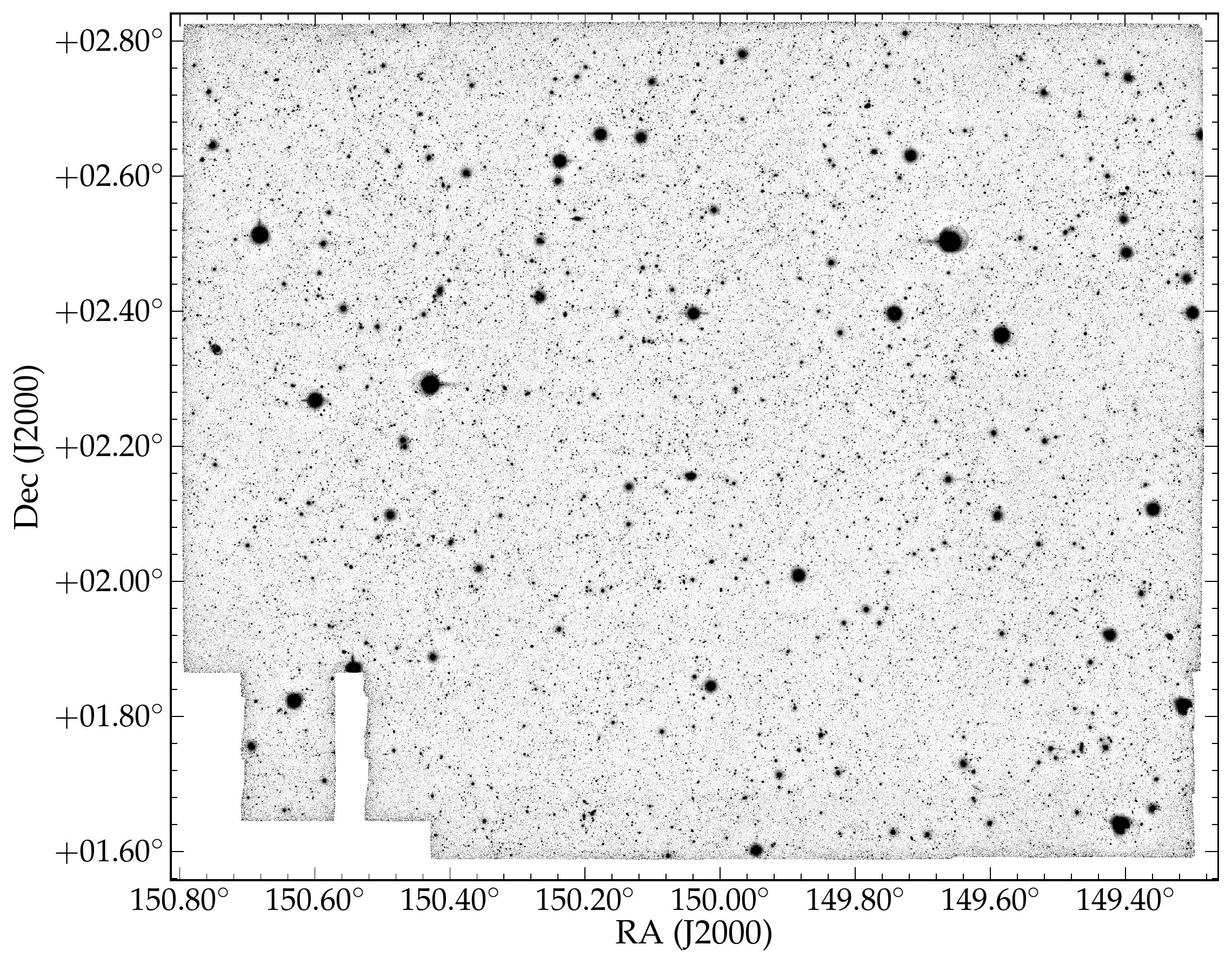}
       \end{center}
       \caption{Full $K_{\rm s}$ mosaic, displayed using a
         logarithmic stretch. The background level is extremely flat,
         and is not perturbed near almost all bright stars. Several
         clusters are visible, corresponding to the many rich
         structures which are present in the COSMOS field.}
   \label{Ksfullframe}
 \end{figure*}

 In the last processing step, the images and weight maps are coadded
 using a modified version of the \texttt{swarp} software
 \citep{Bertin:2002p5282} which permits a combination of images based
 on a clipped sigma estimator; we use a clipping threshold of
 $2.8\sigma$. Before stacking, a small number of images which have
 large photometric extinction or bad astrometric solutions are also
 rejected. For the final stacks, in the five bands, 6520 images were
 used.  Since the size of VIRCAM pixels varies radially as a function
 of distance from the centre of the mosaic, this must be accounted for
 during image co-addition. Bad regions on individual detectors (such
 as half of detector 16, whose pixels suffer from time variable
 quantum efficiency, most notable at shorter wavelengths where the sky
 background is lower) are also masked, which explains the irregular
 appearance in the corner of the stacked
 images. Fig.~\ref{Ksfullframe} shows most of the $K_{\rm s}$ image,
 resampled $2\times2$. The final image is completely free of any
 large-scale gradients, and the background is perfectly flat except
 near the brightest objects in the field.

\begin{figure}
       \includegraphics[width=0.49\textwidth]{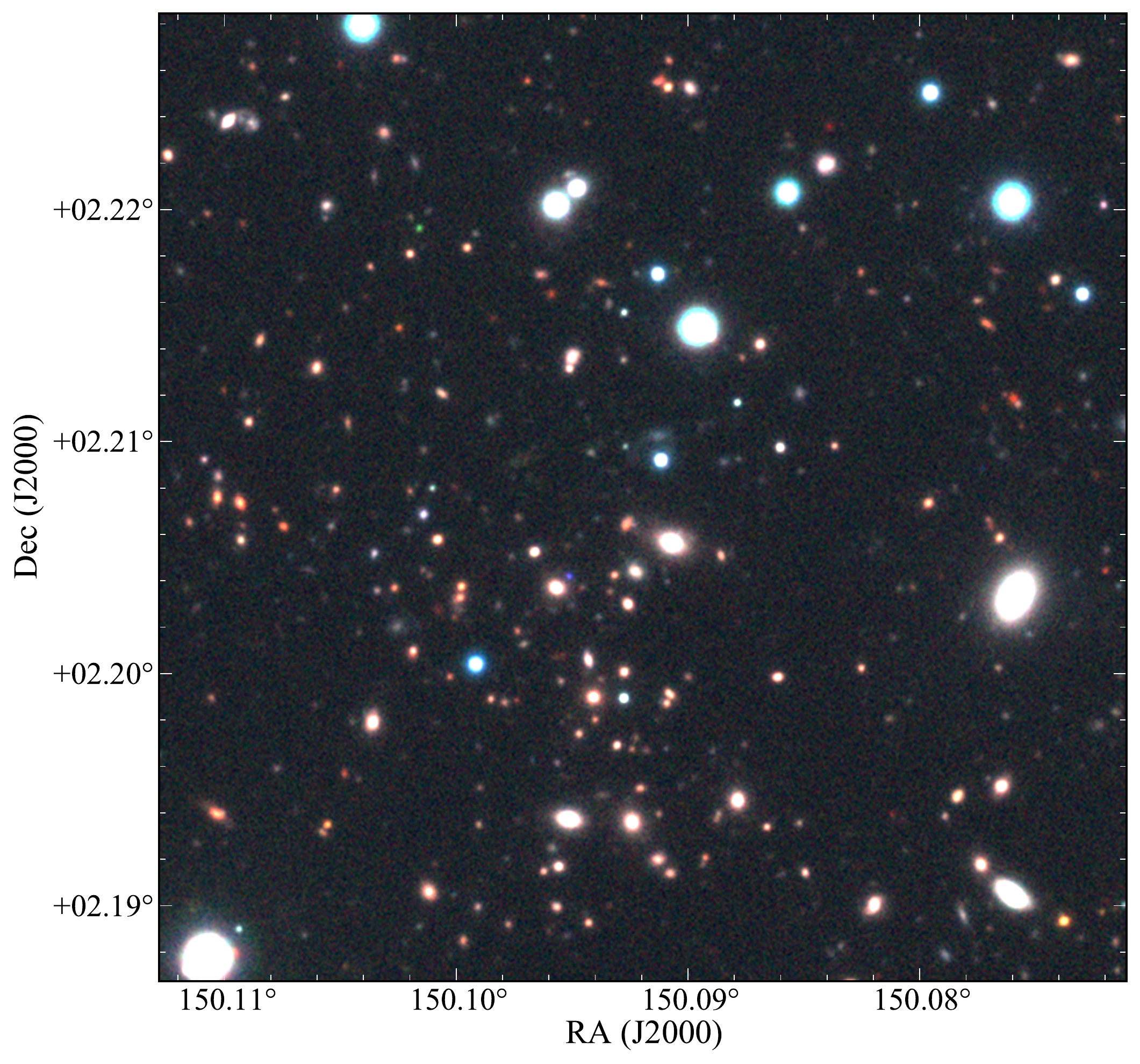}
       \caption{RGB image composed of $K_{\rm s}$, $J$ and $Y$ data
         respectively. The size of this image represents less than
         1/500th of the total area of the field. Sources as faint as
         $K_{\rm s}\sim22$ are easily visible.}
   \label{imageRGB}
 \end{figure}

 In this release, five stacked images and their corresponding weight
 maps are made available for $Y$,$J$,$H$,$K_{\rm s}$ and $NB118$ data
 taken during the first year of public survey operations of the
 UltraVISTA survey. These images have a zero point of 30.0~AB
 magnitudes for an effective exposure time of one second and a pixel
 scale of $0.15\arcsec$/pixel. The weight-maps correspond to
 \texttt{swarp}'s image type \texttt{MAP\_WEIGHT} which correspond to
 maps of relative inverse variance.  Fig.~\ref{imageRGB} shows an RGB
 image composed $K_{\rm s}JY$ images of a small section of the final
 field, illustrating the excellent image quality and depth of our
 final stacks. The bright saturation limit for stellar sources
   in these catalogues is $\sim 14$ magnitudes in $Y$ and 15
   magnitudes in $YJHK_{\rm s}$ bands.

 The images all have a common tangent point, in decimal RA, DEC of
 (1.501163213,2.200973097), corresponding to the tangent point of the
 publicly available IRSA/COSMOS images. Each image (uncompressed) is
 $\sim$ 9Gb in size. This common tangent point and pixel scale means
 that the UltraVISTA survey images are pixel-matched to publicly
 available COSMOS data.

 Finally, to prepare these data products for ingestion in ESO's
 ``phase three'' system, all the image and table headers produced were
 edited to comply with the Phase 3 requirement document, including
 most of the information which is presented here in the FITS header
 keywords. 

 Table~\ref{tab:image-summmary} summarises the principal properties of
 each coadded stack. In each case we report the average seeing over
 the full mosaic, the $95\%$ completeness limit, and the limiting
 magnitude. We also list the typical exposure time per pixel for each
 stack as well as the total on-sky integration time. Summed over all
 filters, this is 55 and 155hrs respectively for the data presented
 here. Seeing on the final stack is characterised using the
 \texttt{PSFex} tool. The average seeing is calculated from a fit to a
 \cite{Moffat:1969p12721} profile. We note that in $Y$ band the PSF
 has slightly broader wings compared to redder bandpasses (with a
 best-fitting Moffat $\beta$ parameter which varies from $\sim2.4$ in
 $Y$ to $\sim3.5$ in $K_{\rm s}$).

 Limiting magnitudes are computed as follows: first,
 \texttt{SEXtractor} is run on each stack using the same detection
 threshold parameters as used for catalogue generation. All pixels
 belonging to objects to this detection limit are flagged. Next, we
 measure fluxes in apertures of diameter $2\arcsec$ over the entire
 mosaic; any aperture which contains object pixels is discarded. The
 limiting magnitude is then simply computed from the standard
 deviation of fluxes measured in these apertures.  Our completeness
 statistics are computed by adding artificial stars to the images with
 average image FWHM and then measuring the fraction which are
 successfully detected with \texttt{SExtractor} using the same
 measurement parameters used for the catalogues.

\begin{table*}
\caption{Characteristics of the stacked images.}
\centering
  \begin{tabular*}{\linewidth}{@{\extracolsep{\fill}} c c c c c c}
    \hline\hline
    Filter & Typical exposure time per pixel & Total exposure time &
    $5\sigma$($2\arcsec$) ($\pm 0.1\mathrm {~mag}$) & 95\% comp. ($\pm 0.1\mathrm {~mag}$)&  seeing
    ($\arcsec$)($\pm 0.1\arcsec$)\\
    \hline
    $Y$ & 42360 &127080 &24.6&  24.2 &  0.82\\
    $J$ & 49720 &149160 &24.4&  24.2 & 0.79 \\
    $H$ & 42520 &127560 &23.9& 24.1 & 0.76 \\
    $K_{\rm s}$ & 39400 &118200 &23.7& 23.8 & 0.75 \\
    NB118 & 23773 &35660 &$22.9\pm0.2$ & 22.6 & 0.75 \\
    \hline
\end{tabular*}
\tablefoot{The seeing is computed from a fit to a
  \cite{Moffat:1969p12721} profile.}
\label{tab:image-summmary}
\end{table*}

\begin{figure}
       \includegraphics[width=0.49\textwidth]{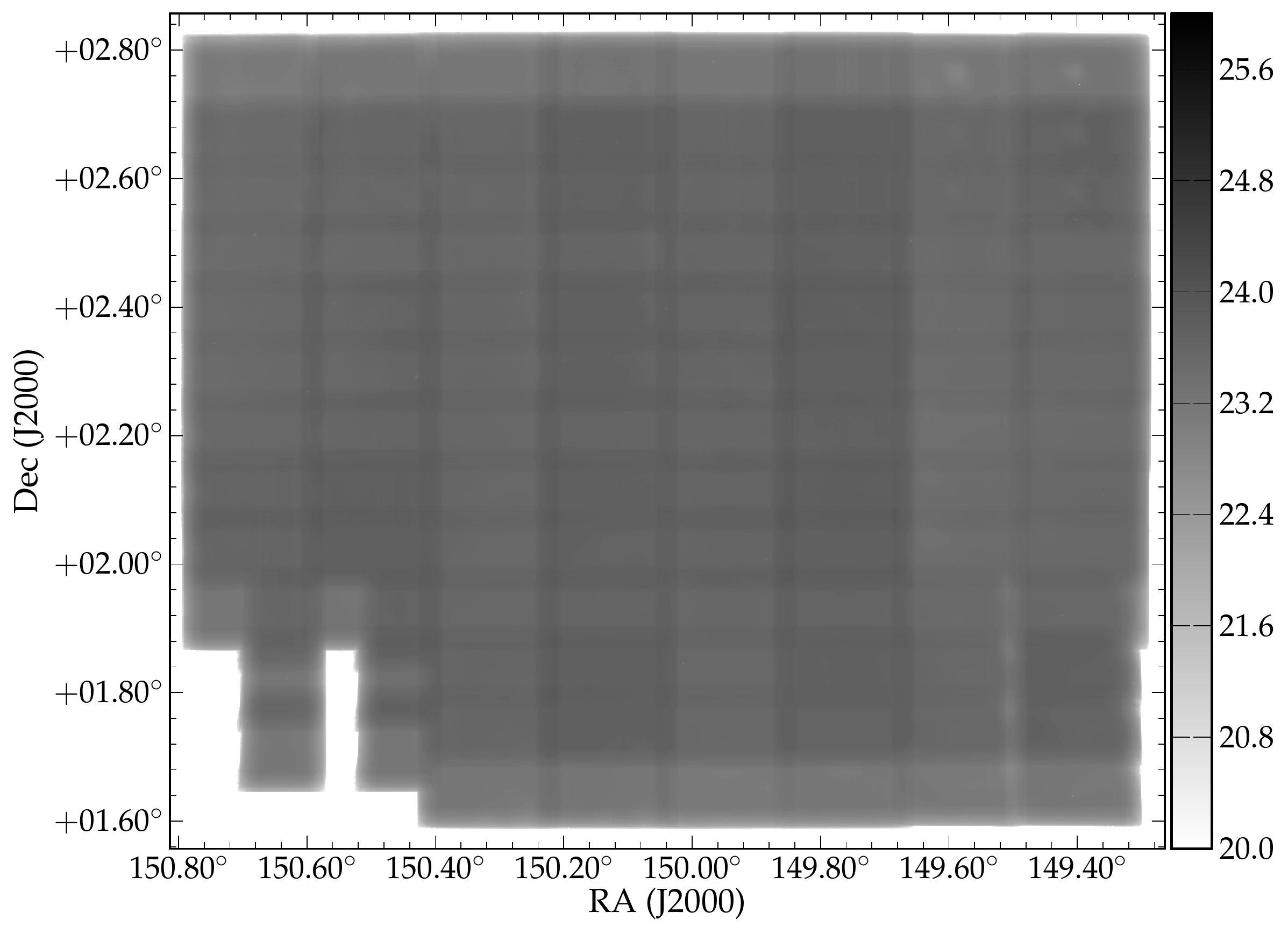}
       \caption{Weight-map for the first-year $K_{\rm s}$-band
         data. The intensity at each pixel has been converted to an
         approximate $5\sigma$ limiting magnitude for an aperture of
         $2\arcsec$ diameter. The strips at the top and bottom of the
         image have half the average exposure time per pixel. }
   \label{comp-map}
 \end{figure}

Fig.~\ref{comp-map} shows the weight-map from the first year of
$K_{\rm s}$ observations described in this paper. The intensity at each
pixel has been converted to an approximate limiting magnitude for a
detection in a $5\sigma, 2\arcsec$ aperture. It is important to note
that our weight map is quite uniform, thanks to our adopted observing
strategy.

 From these stacks, two sets of catalogues are provided at the ESO
 archive: those extracted on individual images, and matched catalogues
 which use the $K_{\rm s}$ band image as a detection image. Aperture
 magnitudes reported in the catalogues are measured in $2\arcsec$ and
 $7.1\arcsec$ diameters respectively. Based on the average stellar
 profiles each of the four broad-band filters, these aperture
 magnitudes can be ``corrected'' to pseudo-total magnitudes by adding
 $\sim -0.35, -0.3,-0.2, -0.2$ magnitudes to $Y,J,H,K_{\rm s}$
 $2\arcsec$ aperture magnitudes. These corrections are not applied to
 the catalogues delivered to the ESO archive but they \textit{are}
 applied to the colour-colour plots shown in
 Section~\ref{sec:colo-colo-diagr}.

\section{Data quality assessment}
\label{sec:data-qual-assessm}

\subsection{Galaxy number counts}
\label{sec:galaxy-number-counts}

\begin{figure}
\resizebox{\hsize}{!}{\includegraphics{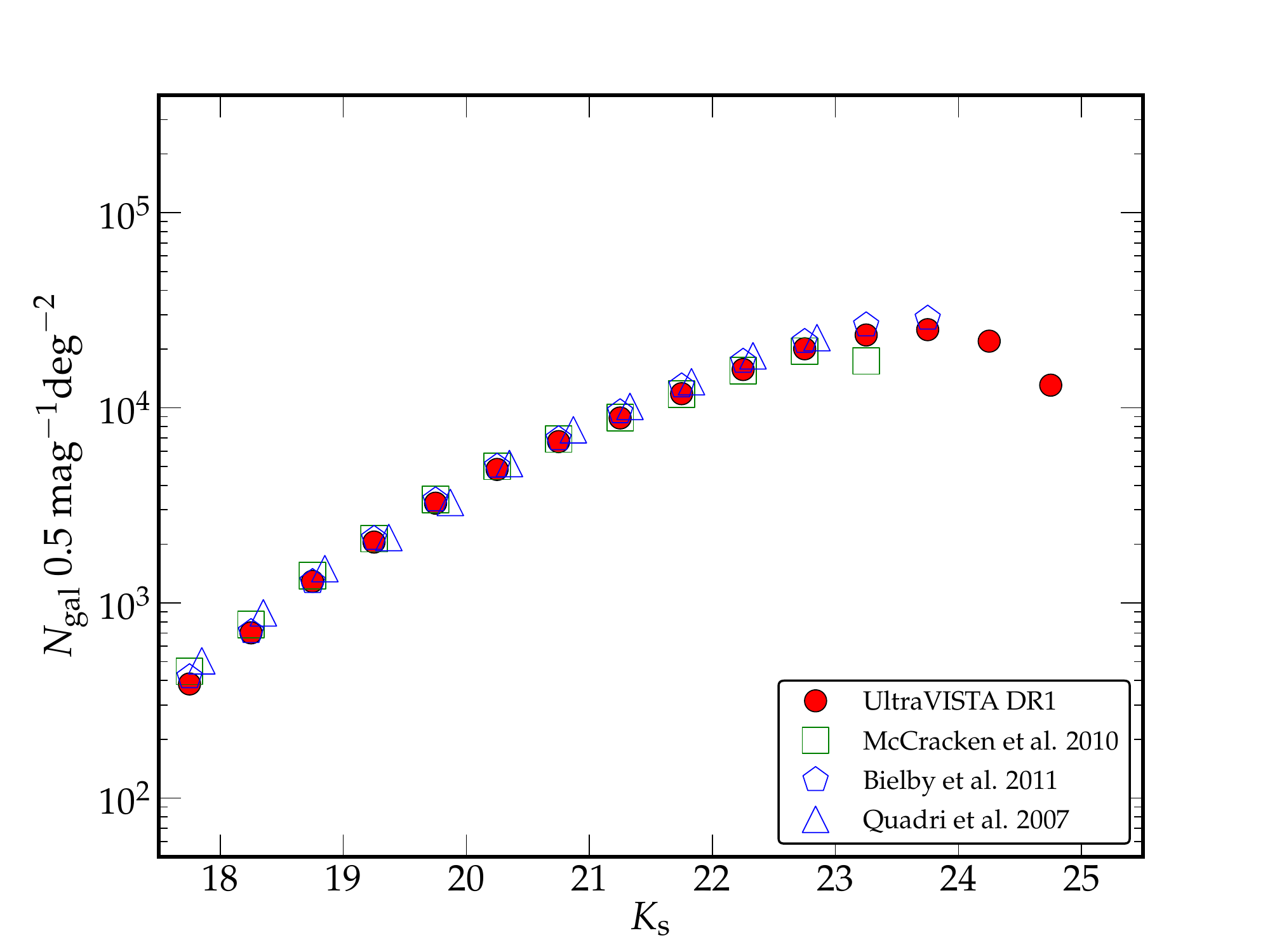}}
\caption{$K_{\rm s}$-selected galaxy number counts for UltraVISTA, in
  addition to some recent wide-field near-infrared surveys. The
  agreement with previous studies is excellent.}
\label{fig:kbandcounts}
\end{figure}

Fig.~\ref{fig:kbandcounts} shows the $K{\rm s}$-band number counts
extracted from our catalogues in comparison with recent literature
measurements, in particular from the wide-area survey ``WIRDS''
carried out using WirCAM at the CFHT \citep{Bielby:2011p12309} and
from COSMOS \citep{McCracken:2010p10723}.  Not surprisingly, our
counts agree well with the existing COSMOS $K_{\rm s}$ counts but also
reach $1\sim$ mag deeper. We are in good agreement with the other,
independent studies covering smaller areas than our work, for example
\citep{Quadri:2007p4415}. 

\subsection{Astrometric comparisons with external catalogues}
\label{sec:comp-with-astrometric}

\begin{figure}
\resizebox{\hsize}{!}{\includegraphics{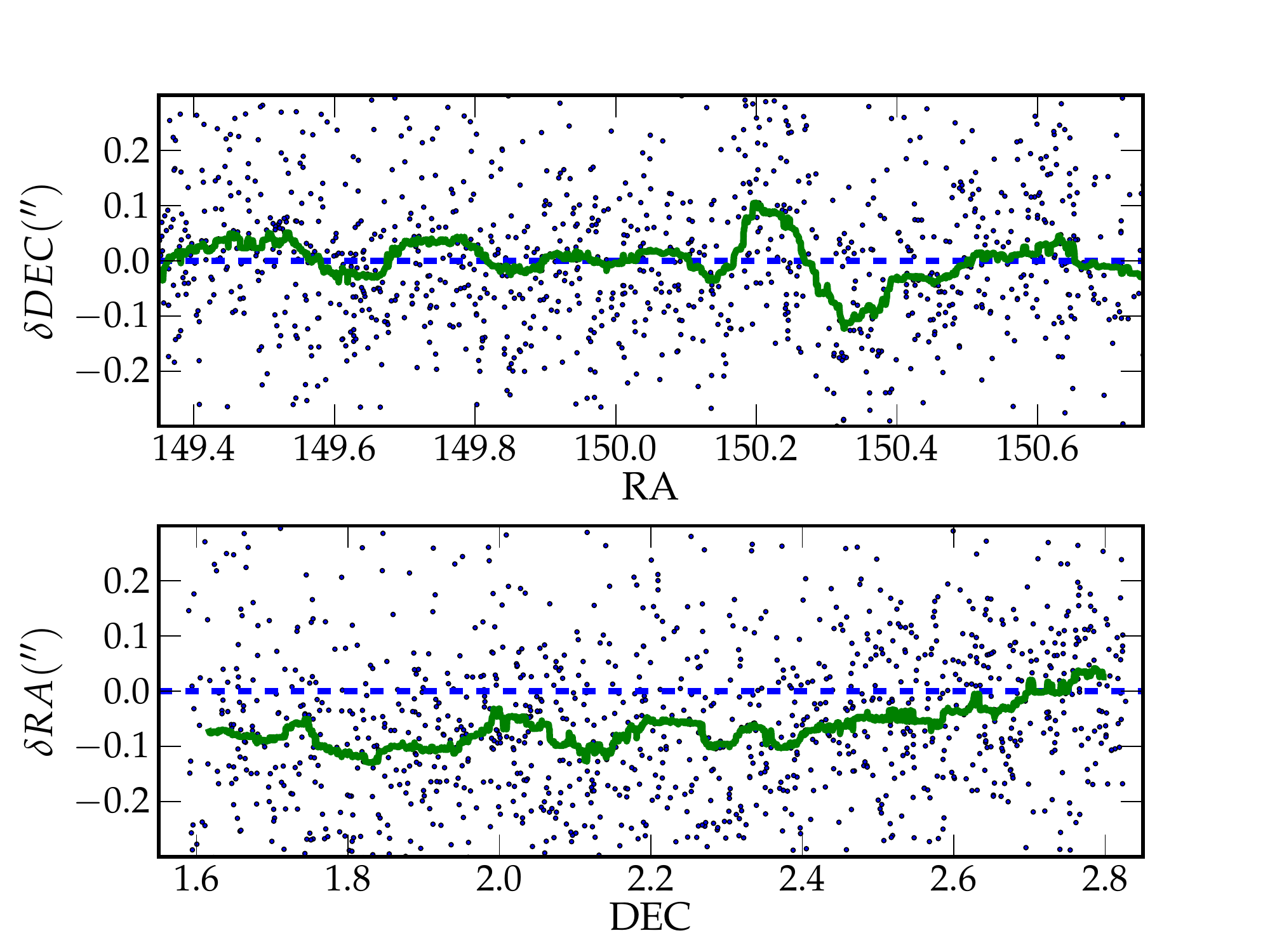}}
\caption{Difference in position, in arcseconds, with respect to
  stars in the 2MASS, as a function of right ascension and declination
  (upper and lower panels respectively); every second point is plotted. The solid line shows a
  running median. The RMS in both axes is $\sim0.15\arcsec$.}
\label{fig:comparison2massastro}
\end{figure}

We compare the positions in right ascension and declination of
point sources in 2MASS with those in our UltraVISTA $K_{\rm s}$
catalogue. This is shown in
Fig.~\ref{fig:comparison2massastro}. Note that, unlike for our
photometric solutions, we do not use 2MASS as our astrometric
reference catalogue but use instead a densely-sampled catalogue from
the COSMOS CFHT $i-$band observations. The absolute astrometric
calibration of COSMOS is derived from VLA 20cm observations
\citep{Schinnerer:2004p12717}, and these positions are known to be
offset slightly with respect to 2MASS~\citep{Capak:2007p267}, which is
indeed what we observe. Our median offsets and $1\sigma$ RMS with
respect to 2MASS is $(0.00,0.14)$ arcsec and $(-0.07,0.15)$ arcsec in
RA and DEC respectively.

\begin{figure}
\resizebox{\hsize}{!}{\includegraphics{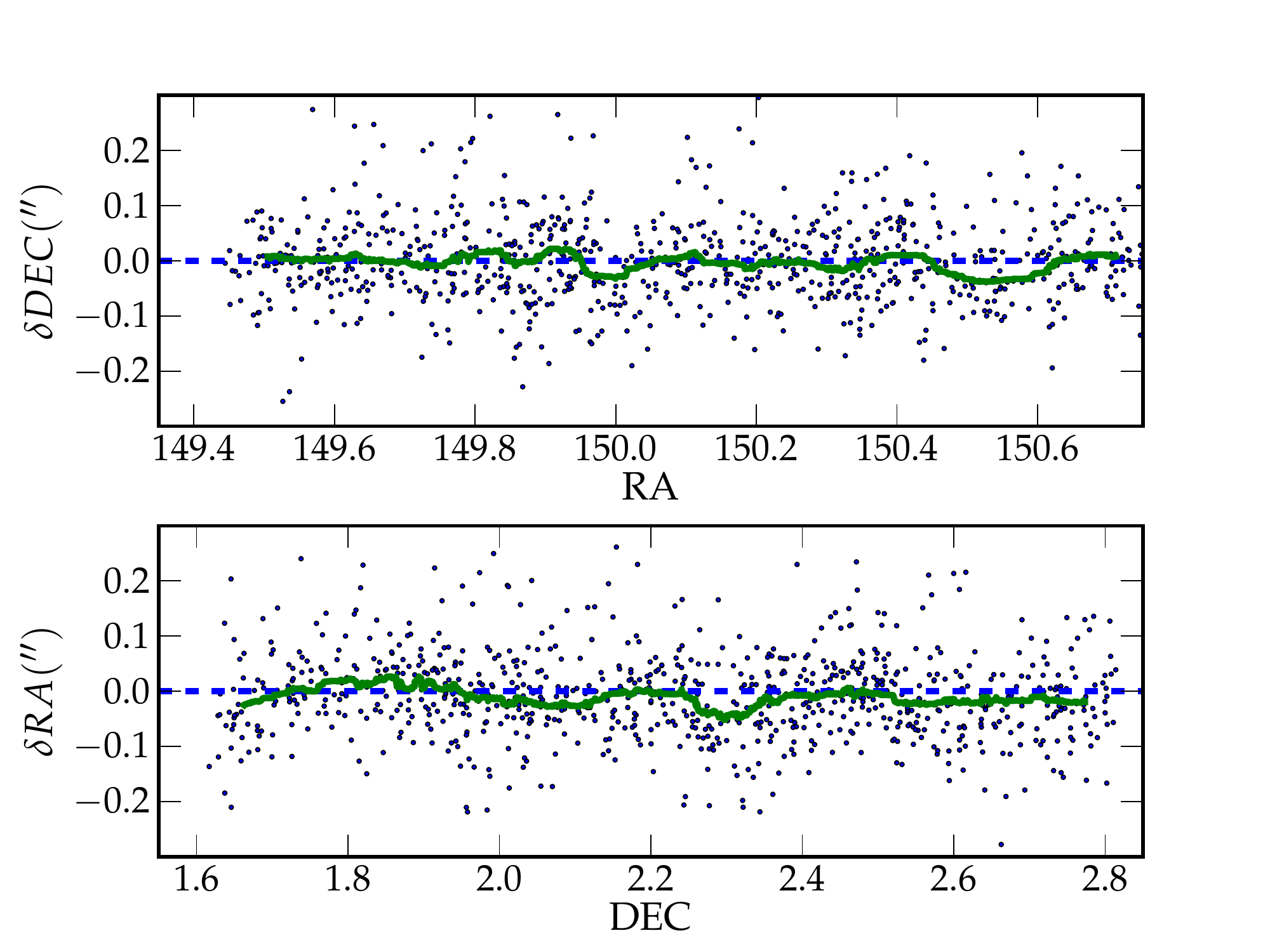}}
\caption{Difference in position in arcseconds for stars between
  the public ACS catalogue of \cite{Leauthaud:2007p5538} and the
  UltraVISTA $K_{\rm s}$ stack; every second point is plotted. The
  solid line shows a running median. For both axes the median
  residuals are $\lessapprox 0.05\arcsec$.}
\label{fig:acs-comparison}
\end{figure}

To verify that our astrometric reference frame is consistent with
COSMOS, we carried out a similar comparison with stars in the COSMOS
ACS catalogue \citep{Leauthaud:2007p5538}; this is shown in
Fig.~\ref{fig:acs-comparison}. In RA and DEC, no offset is
observed. The $1\sigma$ RMS in both directions for stars selected with
$17.0<K_{\rm s}<19.5$ is $\sim 0.08$ arcsec. The internal astrometric
accuracy between different UltraVISTA bands is expected to be of this
order or better, i.e., much better than one $0.15\arcsec$ pixel.

\subsection{Photometric comparisons with external catalogues}
\label{sec:comp-with-cosm}

\begin{figure}
\resizebox{\hsize}{!}{\includegraphics{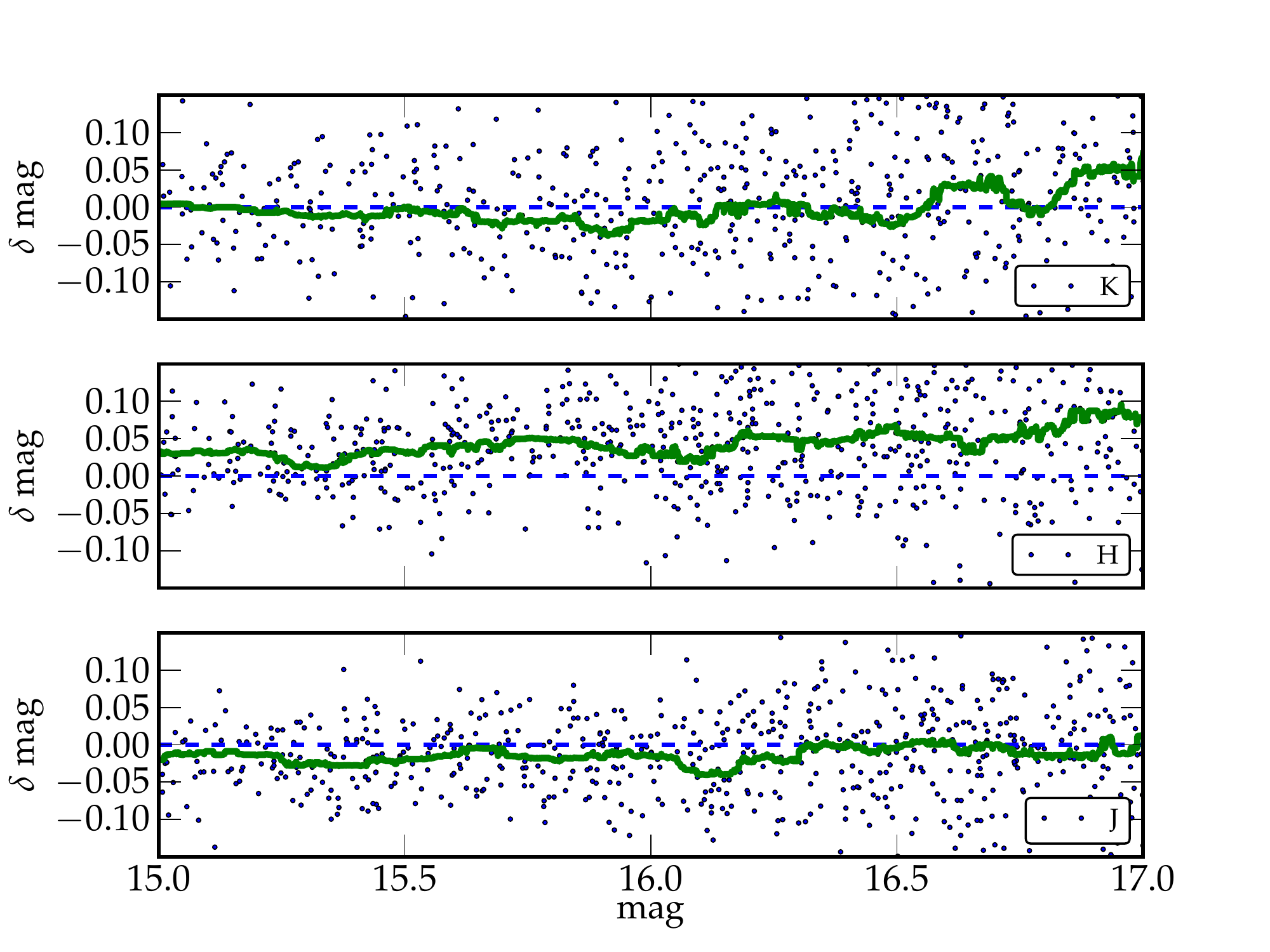}}
\caption{Difference between total $J, H, $ and $K_{\rm s}$
  magnitudes of stars in UltraVISTA with sources in 2MASS. The green
  line corresponds to a running median.}
\label{fig:comparison2mass}
\end{figure}

We compare the total magnitudes of stars in our catalogue
(\texttt{mag\_auto}) with those in the 2MASS all-sky point source
catalogue \citep{Skrutskie:2006p3517}. (Note also that 2MASS is used
for the photometric calibration of the survey by CASU.) Of course, a
significant limitation of this comparison is that the magnitude range
over which sources in UltraVISTA and 2MASS overlap is relatively
small. Nevertheless, the result of this test is shown in
Fig.~\ref{fig:comparison2mass} where we plot UltraVISTA-2MASS
magnitudes for all non-saturated stellar sources and for a total
photometric error in (2MASS and UltraVISTA, summed in quadrature) of
less than 0.2 magnitudes. The thick solid line shows a running median
which is always within 0.05 magnitudes of zero for $15.0<\rm
{mag}<17.0$. There is a slight systematic offset visible in $H$ ($\sim
0.03$) magnitudes; this could be due to incorrectly rescaling our
exposures to slightly non-photometric images or a real offset between
the two different photometric systems.

\begin{figure}
\resizebox{\hsize}{!}{\includegraphics{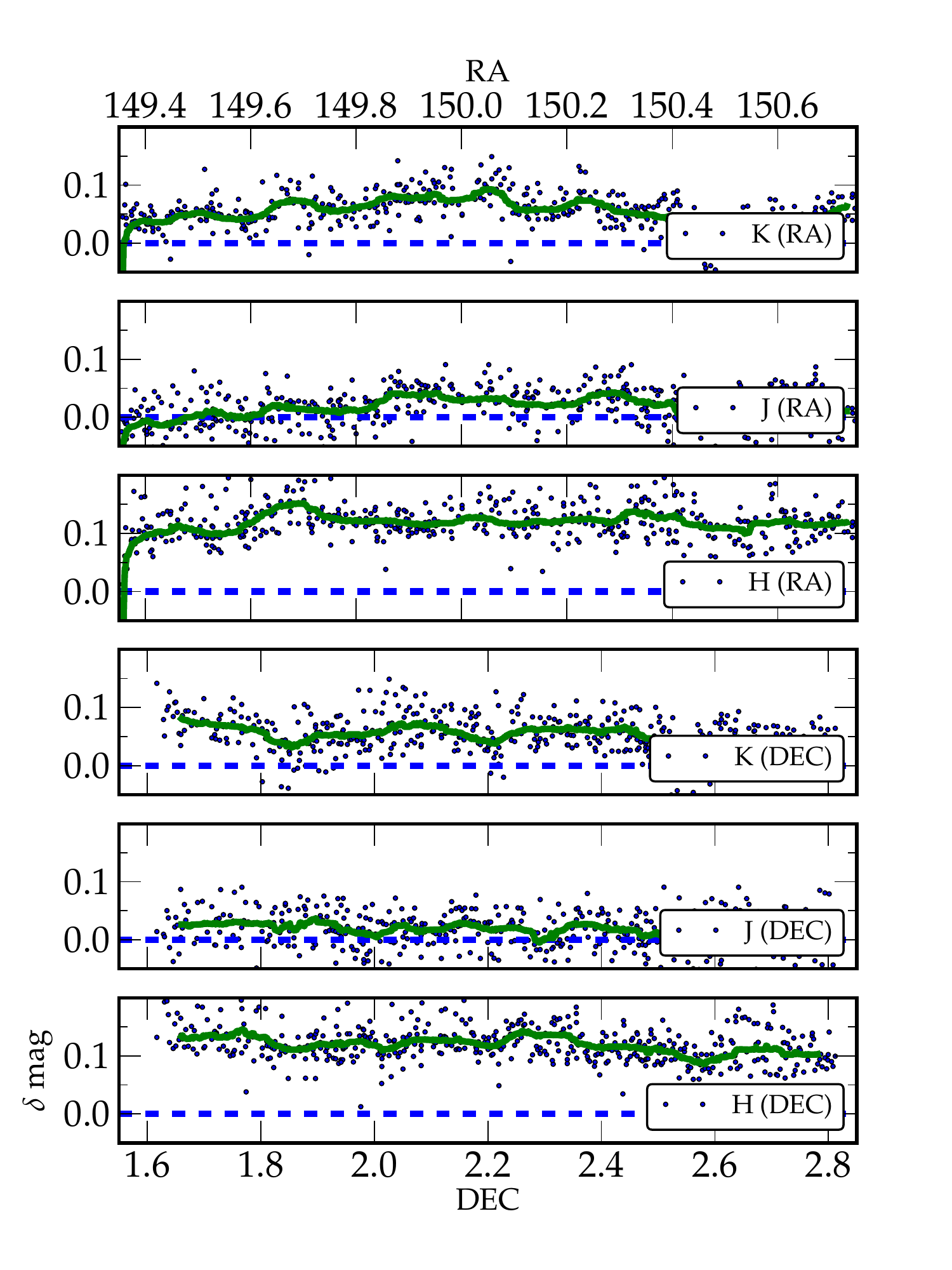}}
\caption{Difference in $J,H$ and $K_{\rm s}$ total magnitudes
  between $BzK$-selected stars with $17<K_{\rm s}<19$ in UltraVISTA
  and COSMOS as a function of right ascension (top three panels) and
  declination (bottom three panels). For clarity, only every fourth
  point is plotted. The thick green line corresponds to a sliding
  median calculated from a window of 100 points. In all cases, the
  differences with the COSMOS photometry is less than 0.1 magnitudes.}
\label{fig:jhkcosmosdiff}
\end{figure}

This Section presents photometric comparisons between UltraVISTA and
COSMOS $JHK_{\rm s}$ measurements. A large amount of near-infrared
observations have already been accumulated on the UltraVISTA field by
the COSMOS team. These consist of $K_{\rm s}$-
\citep{McCracken:2010p10723} and $H$- band observations made with
WIRCam on the CFHT and $J$-band observations made with WFCAM on
UKIRT. In all cases, these observations are shallower than the
first-year UltraVISTA data set presented here. Since our stacks have
the same pixel scale and tangent point as the public COSMOS data, to
make our comparisons we can simply run \texttt{sextractor} in
``dual-image'' mode, choosing as detection image the UltraVISTA
$K_{\rm s}$ image and as measurement images the publicly-available
COSMOS $K_{\rm s}$, $H$ and $J$ stacks. This approach ensures that no
source matching errors are introduced. The results of this comparison
is shown in Fig.~\ref{fig:jhkcosmosdiff}. For test sources we choose
$BzK-$selected stars, as described in the following section.

In Figs.~\ref{fig:jhkcosmosdiff} there is an offset of $\sim0.1-0.05$
magnitudes between UltraVISTA and the publicly-available COSMOS
$HK_{\rm s}$ data. We note that at brighter magnitudes, UltraVISTA
magnitudes are in good agreement with 2MASS, at least for $K_{\rm s}$
and $J$, bands, and for $H$ the offset reported with respect to 2MASS
is smaller than the offset with respect to COSMOS magnitudes. There is
also some evidence in the $K_{\rm s}$ data of a position-dependent
offset. Without a third, equally deep data set, it is hard to know
with certainty the origin of these offsets (especially as the VISTA
and COSMOS data photometric systems are not identical). Furthermore,
examining the magnitude of the offsets with respect to the COSMOS and
UltraVISTA weight-maps they do not seem to be correlated with position
on the focal planes of either instrument (which might be the case if
there was a problem with the photometric calibration on a chip-by-chip
basis). A definitive resolution to this issue awaits more involved
tests, such as photometric redshift comparisons with spectroscopic
data, which will be the subject of a future article.

\subsection{Seeing variation across the mosaics}
\label{sec:seeing-vari-across}

As described above, the final UltraVISTA stack is comprised of six
separate ``pawprints''. At each pawprint the telescope jitter
displacement is less than the separation between the detectors, so no
detectors overlap. In general, each OB typically contains only images
jittered around a single pawprint position, and consequently the
observing conditions, in particular the average seeing is not always
identical pawprint-to-pawprint.  In first-year data presented here, OBs had a
mix of maximum seeing constraint between $0.8\arcsec$ and
$1.0\arcsec$; furthermore there is no minimum seeing cut. A
consequence of this is that when the observations are separated
paw-by-paw, in some filters, there is a variation of around $5\%-10\%$
in average seeing over all 16 detectors from paw-to-paw. In the final
stack, which is the combination of all pawprints, this is visible as bands
of regions of slightly different seeing.

In a future UltraVISTA release we will make available stacks for which
we have carried out a paw-level homogenisation (in which each of the
six pawprints are convolved by a Gaussian to bring them to a common
FWHM). For the moment, we report here that this effect is important
for the $K_{\rm s}$ and $H-$ band stacks. In Fig.~\ref{fig:flux-radius
  plot} we show the seeing, calculated from \texttt{SExtractor}'s
\texttt{FWHM\_WORLD} parameter (which is derived from the isophotal
area of the object at half maximum, and so may not be comparable to
the figures listed in Table~\ref{tab:image-summmary}), as a function
of right ascension and declination. Because of a sequence of pawprints
with significantly better seeing, there is around a $5\%$ variation in
seeing as a function of right ascension.

\begin{figure}
\resizebox{\hsize}{!}{\includegraphics{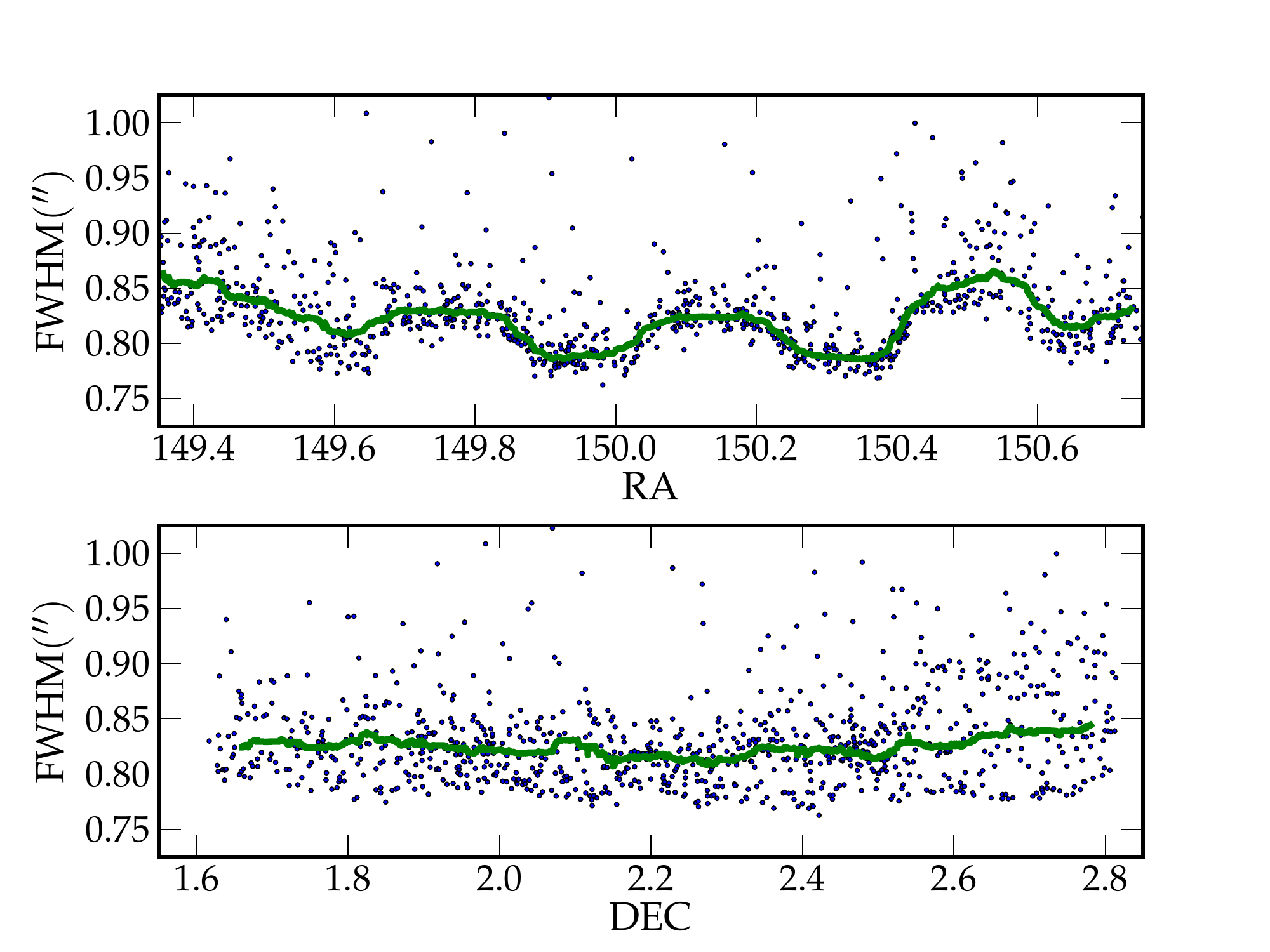}}
\caption{Seeing FWHM for stars (corresponding to \texttt{SEXtractor}'s
  \texttt{FWHM\_WORLD} parameter) selected in the $BzK$ diagram, as a
  function of RA and DEC, in the $K_{\rm s}$ stack (every fourth point
  is plotted). As before, the solid green line corresponds to a
  running median. The seeing variations are small, of order $\sim
  0.05\arcsec$, and vary principally as a function of RA. }
\label{fig:flux-radius plot}
\end{figure}

\subsection{Colour-magnitude and colour-colour diagrams}
\label{sec:colo-colo-diagr}

\begin{figure}
\resizebox{\hsize}{!}{\includegraphics{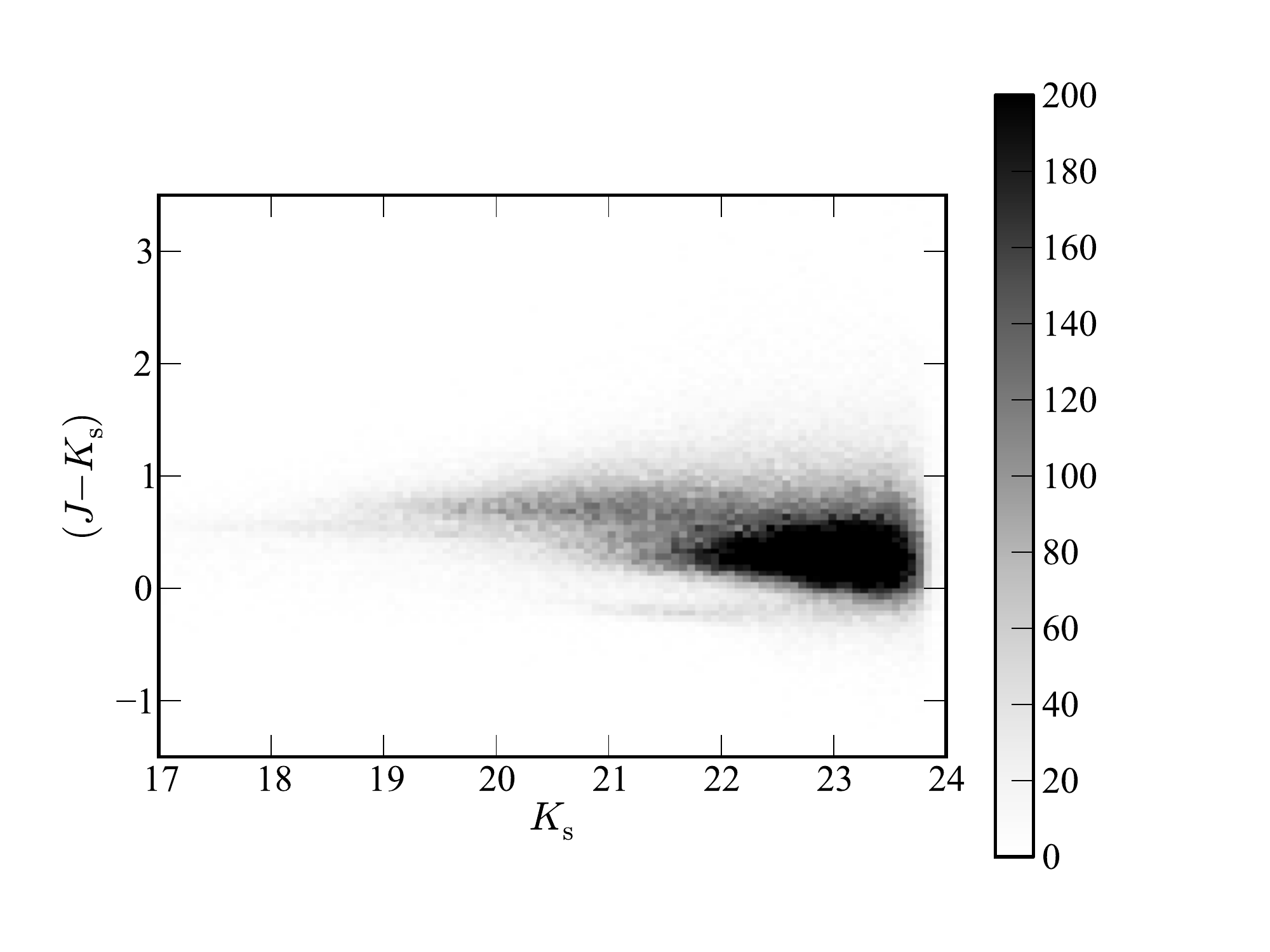}}
\caption{Two-dimensional histogram showing $(J-K_{\rm s})$ corrected
  aperture colour as a function of $K_{\rm s}$ total magnitude; the
  grey level at each bin in magnitude-colour space corresponds to the
  surface density of objects. The narrow ridge clearly visible at
  $(J-K_{\rm s})\sim -0.2$ corresponds to the location of stellar
  sources.}
\label{fig:kjk_greyscale}
\end{figure}

The large number of sources in our catalogues combined with our
excellent seeing and high signal-to-noise means that we can
investigate in detail the distribution of objects in colour-colour
space. In Fig.~\ref{fig:kjk_greyscale} we plot the $(J-K_{\rm s})$ vs
$K_{\rm s}$ distribution of sources in our $K_{\rm s}$ selected
catalogue. The stellar locus is clearly visible as a narrow ridge of
constant $(J-K_{\rm s})$ colour (which one can confirm by overplotting on this
diagram the location of stars identified in the ACS catalogue). 

Next, we consider the distribution of objects in optical and
near-infrared colour-colour space, turning first to the ``$BzK$''
diagram as this allows us to cleanly separate stars and galaxies. We
use the publicly-available COSMOS $B$ and $z$ Subaru images
\citep{Capak:2007p267} and transform the $B$ and $z$ magnitudes in
each catalogue following the recipes in \cite{McCracken:2010p10723} to
bring our system to the ``$BzK$'' system defined in
\cite{Daddi:2004p76}.


\begin{figure}
\resizebox{\hsize}{!}{\includegraphics{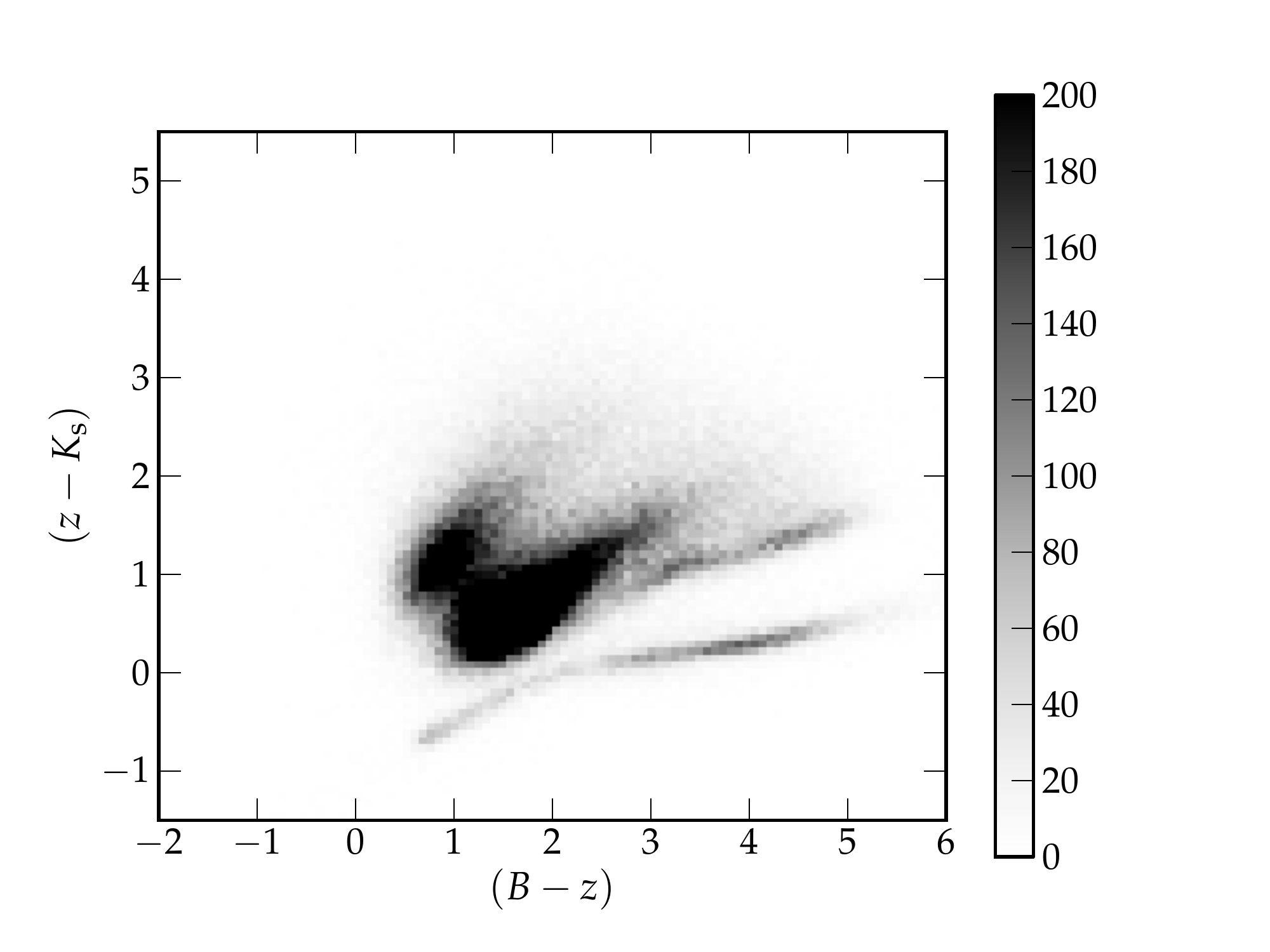}}
\caption{Two-dimensional histogram of $(B-z)$ vs $(z-K_{\rm s})$
  corrected aperture colour for UltraVISTA. All sources detected to a
  $5\sigma$ limit in $K_{\rm s}$ auto magnitudes are shown. The
  stellar locus is clearly visible as a ridge at blue $(z-K_{\rm s})$
  colour.}
\label{fig:color-colour_bz_zk}
\end{figure}

The result is shown in Fig.~\ref{fig:color-colour_bz_zk} as a
two-dimensional grey-scale histogram; in this diagram and all
subsequent diagrams we show all objects detected to $5\sigma$ in
$K_{\rm s}$ band aperture magnitude. Several interesting features are
clearly visible in this diagram: firstly the stellar locus, which is
apparent as the long ``ridge'' feature which is relatively blue in
$(z-K_{\rm s})$; secondly, almost parallel to the stellar locus but redder in
$(B-z)$ is a second ``ridge'' which is comprised mainly of
lower-redshift passive galaxies
\citep{Lane:2007p295,Bielby:2011p12309}. Thirdly, the division between
lower-redshift normal and star-forming galaxies (the ``sBzK'' galaxies
of \cite{Daddi:2004p76}) is clear.

\begin{figure}
\resizebox{\hsize}{!}{\includegraphics{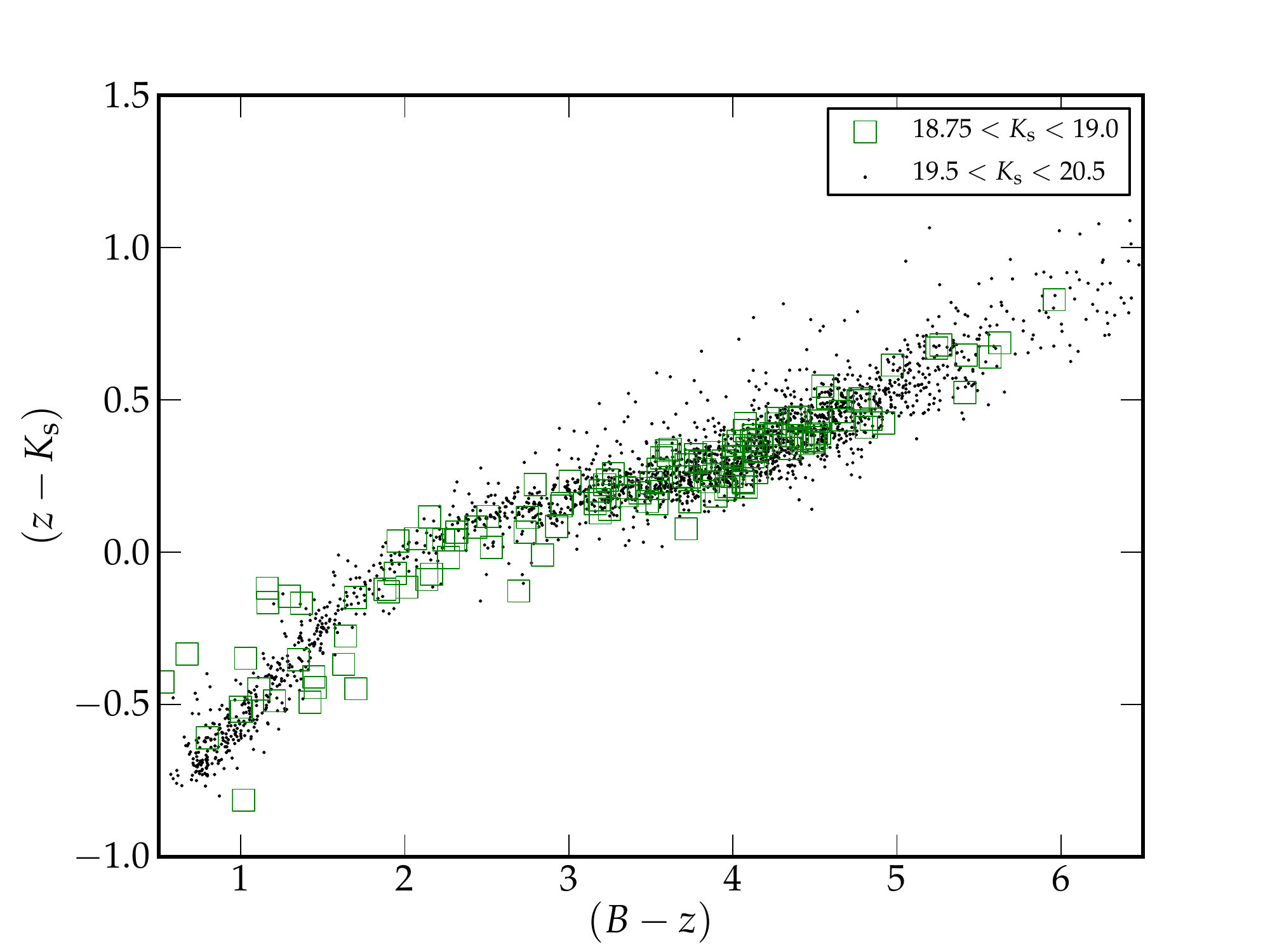}}
\caption{Stellar locus for bright and faint stars in UltraVISTA (shown
  as points and dots respectively) in the $(B-z)$ vs $(z-K)$ corrected
  aperture colour-colour plane. Bright and faint stars occupy the same
  location in colour-colour space.}
\label{fig:color-colour_bzk_zoom}
\end{figure}

In Fig.~\ref{fig:color-colour_bzk_zoom} we show a magnified view of
the stellar locus in the $BzK$ diagram, and we show both bright and
faint stars. The position of the stellar locus does not depend on the
magnitude limit, which demonstrates that there are no
magnitude-dependent effects present in our data which could arise if
there were issues related to an incorrect sky-subtraction.

\begin{figure}
\resizebox{\hsize}{!}{\includegraphics{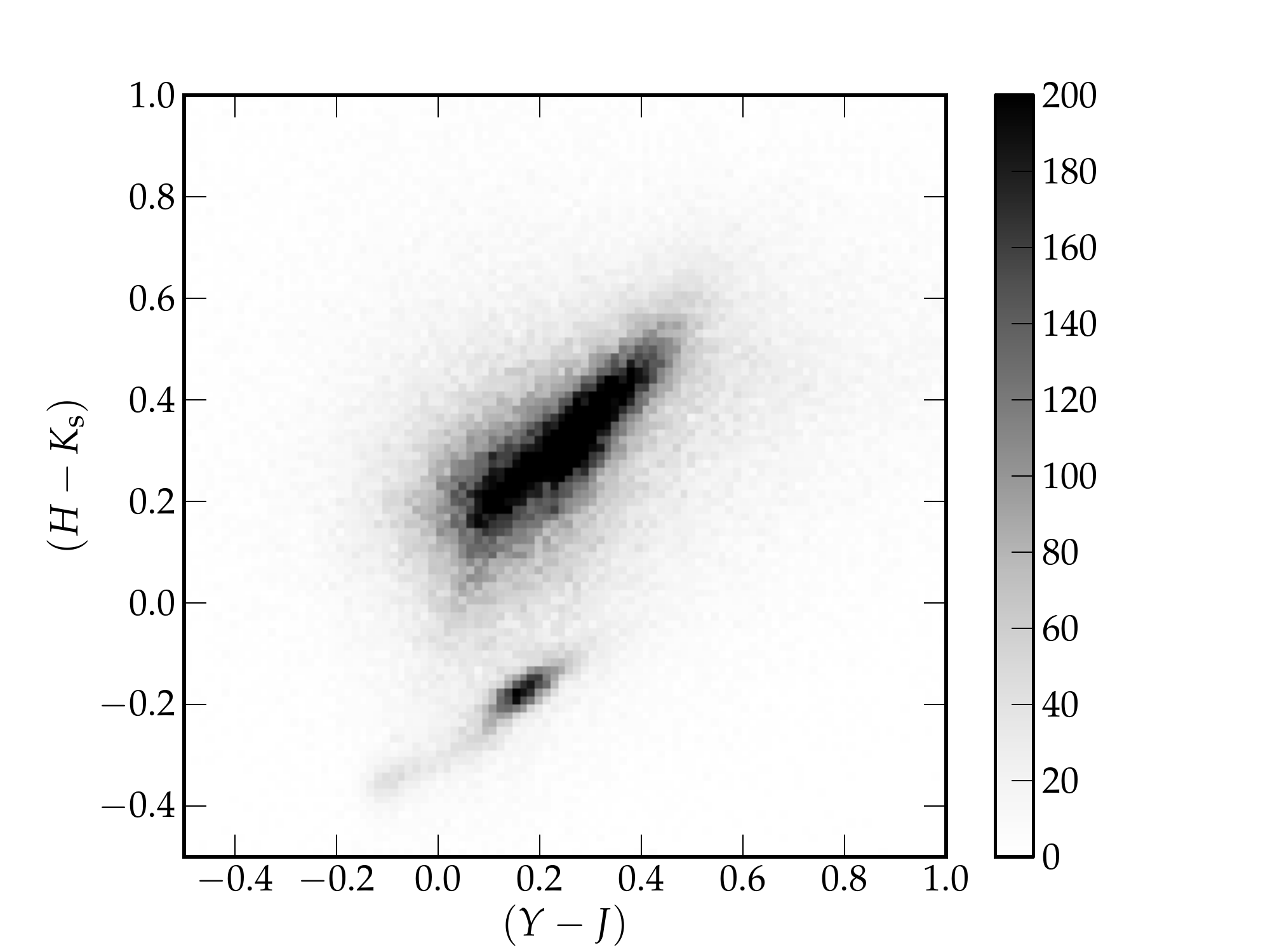}}
\caption{Two-dimensional corrected aperture colour-colour $(H-K_{\rm
    s})$ vs $(Y-J)$ histogram for all sources with total magnitude
  $16.0<~K_{\rm s}<~23$. Stars and galaxies are cleanly separated. The
  stellar locus corresponds to the 'bump' visible at $0.2,0.1$ in
  $(Y-J)$ vs $(H-K_{\rm s})$.}
\label{fig:color-colour_hk_yj}
\end{figure}

We also consider the distribution of galaxies in the purely
near-infrared colour-colour space $(H-K_{\rm s})$ vs $(Y-J)$, shown in
Fig.~\ref{fig:color-colour_hk_yj}. Again, stars and galaxies are
cleanly separated.

\section{Summary and conclusions}
\label{sec:summary}

In this paper we have described the first public UltraVISTA data
release. This data set comprises five high-quality image stacks
representing a unique combination of depth and areal coverage at
near-infrared wavelengths.  Our stacked images reach $5\sigma$ depths
in aperture of $2\arcsec$ diameter of $\sim 25$ in $Y$ and $\sim 24$
in $JHK_{\rm s}$ bands. Furthermore, it is worth noting that these
depths are in agreement with the expected sensitivity of the telescope
at the time of writing the original UltraVISTA survey proposal. To
these limits, our $K_{\rm s}$ catalogue contains 216,268 sources. The
$1\sigma$ astrometric RMS in right ascension and declination for stars
selected with $17.0<K_{\rm s}<19.5$ is $\sim 0.08$~arcsec in
comparison to the publicly-available COSMOS ACS catalogues. Each of
the stacks has sub-arcsecond seeing and the FWHM variation over the
images is less than 5\% in most bands. Our number counts and
photometric calibration are in good agreement with previous studies. 

The images and catalogues described here are publicly available from
the ESO
\fnurl{archive}{http://www.eso.org/sci/observing/phase3/data_releases/ultravista_dr1.html}.

At the present time of writing (April 2012), a further 250 hours of
UltraVISTA observations have been completed. We intend to deliver
regular releases of UltraVista data products as the observations
proceed towards the total 1800h of observation time allocated to the
project.

\begin{acknowledgements}
  H.~J. McCracken acknowledges the use of TERAPIX computing
  facilities. This research has made use of the VizieR catalogue
  access tool provided by the CDS, Strasbourg, France. This research
  was supported by ANR grant ``ANR-07-BLAN-0228''. JPUF and BMJ
  acknowledges support from the ERC-StG grant EGGS-278202. The Dark
  Cosmology Centre is funded by the Danish National Research
  Foundation. OLF, CSJ, LT acknowledge support from the ERC advanced
  grant ERC-2010-AdG-268107. JH acknowledges support from NWO. JSD
  acknowledges the support of the Royal Society via a Wolfson Research
  Merit award, and also the support of the European Research Council
  via the award of an Advanced Grant. The UltraVISTA team would like
  to thank ESO staff for scheduling and making the UltraVISTA
  observations, and the Cambridge Astronomy Survey Unit for providing
  us with pre-preprocessed UltraVISTA images. E. Bertin is thanked
  for useful discussions concerning the data reductions presented here. 
\end{acknowledgements}


\end{document}